\documentclass[12pt,preprint]{aastex}

\usepackage{lscape}
\usepackage{natbib}
\usepackage{enumerate}

\begin{document}

\title{Exploring Halo Substructure with Giant Stars: The Nature of the Triangulum-Andromeda Stellar Features}

\author{Allyson A. Sheffield\altaffilmark{1}, Kathryn V. Johnston\altaffilmark{1}, Steven R. Majewski\altaffilmark{2}, Guillermo Damke\altaffilmark{2}, Whitney Richardson\altaffilmark{2}, Rachael Beaton\altaffilmark{2}, Helio J. Rocha-Pinto\altaffilmark{3}}

\altaffiltext{1}{Department of Astronomy, Columbia University, Mail Code 5246,
New York, NY 10027 (asheffield, kvj@astro.columbia.edu)}

\altaffiltext{2}{Department of Astronomy, University of Virginia, P.O. Box 400325,
Charlottesville, VA 22904 (srm4n, wwr2u, rlb9n@virginia.edu)}

\altaffiltext{3}{Observat\'orio do Valongo, Universidade Federal do Rio de Janeiro, Rio de Janeiro, Brazil (helio@astro.ufrj.br)}

\begin{abstract}
As large-scale stellar surveys have become available over the past decade, the ability to detect and characterize substructures in the Galaxy has increased dramatically. These surveys have revealed the Triangulum-Andromeda (TriAnd) region to be rich with substructure: along with an extension of the Galactic Anticenter Stellar Structure (GASS; also known as the Monoceros system) at distances of $\sim$ 10 kpc a number of features have been detected in this part of the sky in the distance range 20-30 kpc, and the relation of these features to each other -- if any -- remains unclear.  This complex situation motivates this re-examination of the region with a photometric and spectroscopic survey of M giants. 
An exploration using 2MASS photometry reveals not only the faint sequence in M giants detected by Rocha-Pinto et al. (2004) spanning the range $100^{\circ}<l<160^{\circ}$ and $-50^{\circ}<b<-15^{\circ}$ but, in addition, a second, brighter and more densely populated M giant sequence (distinct from GASS). These two sequences are likely associated with the two distinct main-sequences discovered (and labeled TriAnd1 and TriAnd2) by Martin et al. (2007) in an optical survey in the direction of M31, where TriAnd2 is the optical counterpart of the fainter RGB/AGB sequence of Rocha-Pinto et al. (2004). Here, the age, distance, and metallicity ranges for TriAnd1 and TriAnd2 are estimated by simultaneously fitting isochrones to the 2MASS RGB tracks and the optical MS/MSTO features.  
The two populations are clearly distinct in age and distance: the brighter sequence (TriAnd1) is younger (6-10 Gyr) and closer (distance of $\sim$ 15-21 kpc), while the fainter sequence (TriAnd2) is older (10-12 Gyr) and is at an estimated distance of $\sim$ 24-32 kpc. The results also suggest a slight offset in metallicity. However, spectroscopic results reveal trends in radial velocity (averages and dispersions) as a function of Galactic longitude that are \emph{identical} for TriAnd1 and TriAnd2.  A comparison with simulations demonstrates that the differences and similarities between TriAnd1 and TriAnd2 can simultaneously be explained if they represent debris originating from the disruption of the same dwarf galaxy, but torn off during two distinct pericentric passages.
\end{abstract}

\section{Introduction}
\begin{figure}
\plotone{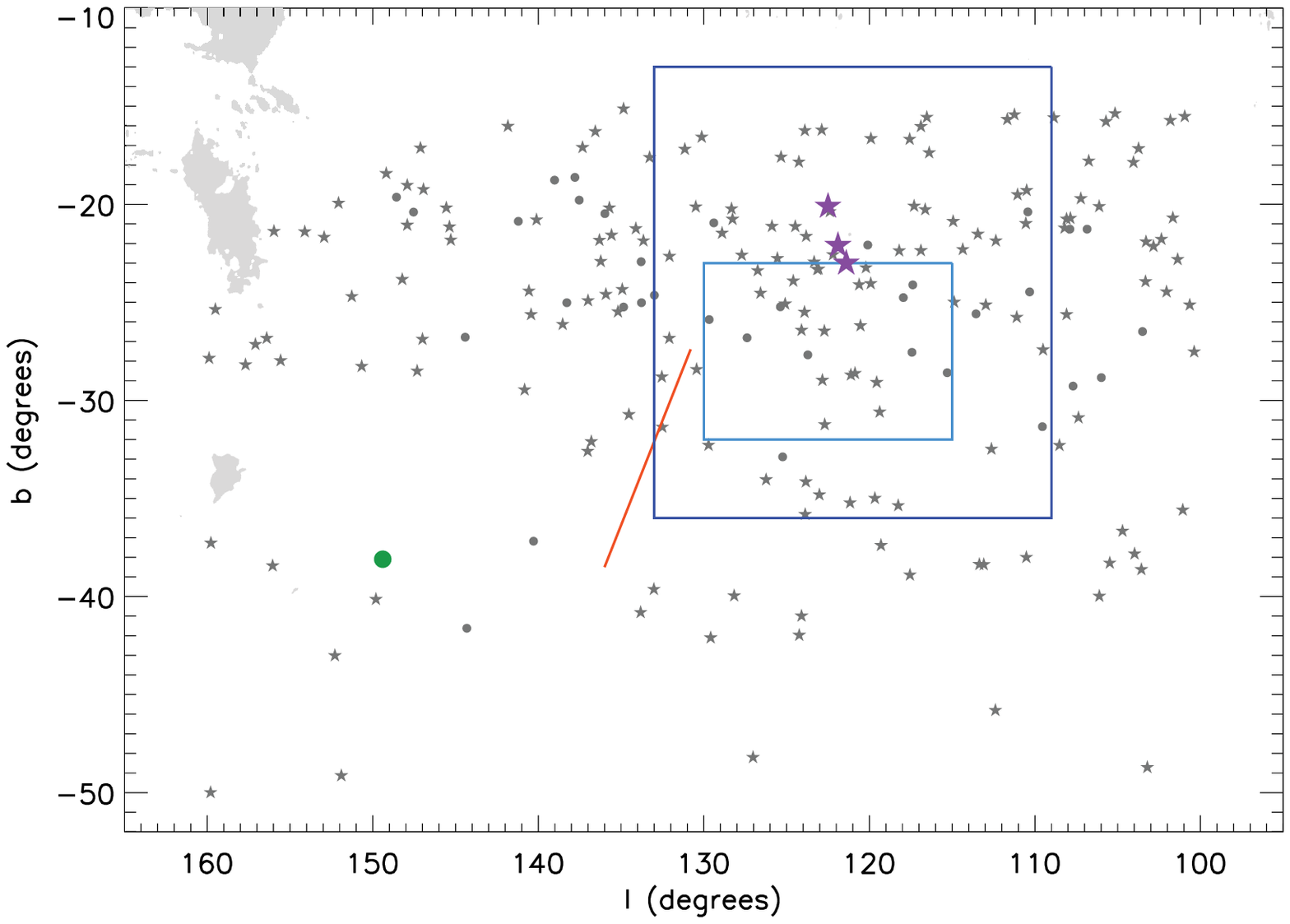}
\caption{\label{subs}Other substructure detections inside our survey region (our program stars are shown by the grey points, with the RP04 TriAnd stars shown as grey circles). The Triangulum Stream \citep{bonaca12} is shown as the orange line; the rough region covered by MegaCam \citep{martin07} is indicated by the cyan box and that for the PAndAS region \citep{martin14} shown by the blue box; the 13 halo stars detected by \citet{deason13} are indicated by the three purple stars, one for each HST pointing; and Segue 2 \citep{belokurov09} is shown as the green circle. The filled light grey regions correspond to areas on the sky with $E(B-V) > 0.555$.}
\end{figure}

Studies of the Milky Way's stellar halo in the region around Triangulum-Andromeda have revealed a profusion of substructures at distances of roughly $\sim$ 20-30 kpc (Majewski et al. 2004; Rocha-Pinto et al. 2004; Martin et al. 2007; Bonaca et al. 2012; Deason et al. 2013; Martin et al. 2013; Martin et al. 2014).  The approximate locations of these detections are shown in Figure \ref{subs}. Initial detections of substructure in this region were made contemporaneously by \citet{maj04} -- by isolating foreground Milky Way dwarfs in a deep photometric survey of M31 that reveal an intervening main sequence (MS) -- and by Rocha-Pinto et al. (2004 -- hereafter RP04),  using M giants from the 2MASS catalog to identify a ``cloud-like'' spatial overdensity (the M giants associated with TriAnd from RP04 are shown as the light grey filled circles on Figure \ref{subs}).  RP04 derived a mean metallicity of $-1.2$ for the TriAnd Cloud and a distance range of 20-30 kpc, and they found a trend in the radial velocities in the Galactic Standard of Rest frame ($v_{\rm GSR}$) as a function of Galactic longitude such that there is a negative gradient in the direction towards the Galactic Anti-center.

Using data from the MegaCam Survey, \citet{martin07} subsequently analyzed 76 deg$^{2}$ in a region southeast of M31 (encompassed by the RP04 TriAnd region --- see the cyan rectangular region in Figure \ref{subs}) and detected two MSs -- referred to here as TriAnd1 and TriAnd2 -- in a deep $(g-i,i)_0$ CMD, and separated by $\sim$ 0.8 magnitudes in the $i$-band.
\citet{martin07} assumed that their brighter sequence (TriAnd1) was associated with RP04's TriAnd Cloud, and, adopting a single isochrone with [Fe/H]=$-1.3$ and an age of 10 Gyr, estimated the TriAnd1/TriAnd2 distances to be 22 kpc and 28 kpc respectively. \citet{martin14} updated their Megacam photometric survey toward the TriAnd region with deep $g$ and $i$ photometry covering 360 deg$^{2}$ from the The Panoramic Andromeda Archaeological Survey (PAndAS) and found the region to be highly substructured, hosting a network of overlying streams and clouds. The approximate region for PAndAS is shown as the blue rectangle in Figure \ref{subs}. 

A very thin stream (width of 75 pc), dubbed ``the Triangulum Stream,'' was detected photometrically in the SDSS DR8 \citep{bonaca12}.  The Triangulum Stream is at an estimated distance of 26 kpc and falls near the center of the TriAnd spatial region identified by RP04 (the feature is shown as the orange line in Figure \ref{subs}).  A separate analysis of the transverse motions of 13 Milky Way halo stars \citep{deason13} (the three purple filled stars in Figure \ref{subs} -- one for each HST pointing) suggested that they might be distributed in a shell-like distribution at a mean distance of 25 kpc. Finally, the faint, dark-matter-dominated dwarf galaxy Segue 2 has also recently been detected in the SDSS \citep{belokurov09}; using four blue horizontal branch stars in Segue 2 (the location is indicated by the green filled circle in Figure \ref{subs}), the distance is estimated by \citet{belokurov09} to be 35 kpc.  

Another prominent feature in this general region of the sky is the Galactic Anticenter Stellar Structure \citep[GASS, also known as ``the Monoceros Ring'', see][]{newberg02,crane03,rp03}. The Megacam and PAndAS $(g-i,i)_0$ CMDs \citep{martin07,martin14} also show the Main Sequence for GASS. The nature of the GASS system is still debated \citep[see, e.g.,][]{li12,slater14}: There are N-body simulations that attempt to fit GASS with tidal debris from an accreted dwarf satellite \citep{penarrubia05}, but there are also claims that GASS results from a disk disturbance \citep{momany06,younger08}.
Moreover, there are apparent continuities in the radial velocities of stars in the GASS and TriAnd features (RP04), leading to the possibility that the two features may be related and be part of the same system \citep{penarrubia05}. High-resolution echelle spectra of stars in both features \citep{chou11}, however, show that the chemical patterns and distances are distinct, which casts doubt on their association: With isochrone fitting, \citet{chou10,chou11} find distances of $\sim$ 12 kpc for the GASS M giants and $\sim$ 22 kpc for the TriAnd M giants.  

Clearly, the nature of and association between the various TriAnd stellar overdensities remains unresolved.  
For example, RP04 speculated that the feature they found could be debris from the tidal disruption of a dwarf galaxy. 
Their survey revealed associated stars covering a remarkable area, spanning nearly 2000 deg$^{2}$ on the sky.
However, the true extent and morphology of their feature remains unclear due to the obscuring foreground of Galactic dust.
The current detections could represent just a small piece of a much longer and broader continuous {\it stream} of stars from the destruction of a satellite on a mildly eccentric orbit, or they could be indicative of a distinct debris {\it cloud} formed from a satellite disrupting on a much more eccentric orbit with its debris collecting at orbital apocenters \citep[the internal view of delicate shell features seen around other galaxies; see][]{johnston08}.
The latter suggestion is particularly intriguing given the more recent work by \citet{deason13} that hint at a cold shell of main-sequence stars at similar distances.
While shells around external galaxies have been studied extensively, clouds around the Milky Way have so far received little attention.
Stars in debris structures associated with clouds can briefly pass close to the Galactic center as they rapidly flow through the pericenters of their eccentric orbits. Thus, connecting debris in the inner Galaxy to the more distant members of the same debris structures in clouds at the apocenters is a promising way of probing the radial density profile of dark matter in the Galaxy \citep{johnston12}.  
These potentially valuable gains in understanding properties of the Galactic halo at both local and intermediate scales motivated us to further explore the nature of the TriAnd stellar features.

In this work, we present an expanded survey of M giants in the TriAnd region, building on the M giant study of \citet{rp04}.  This paper is organized as follows: \S\ref{phot} explores the photometric properties of late-type 2MASS stars in the TriAnd region; in \S\ref{data}, we describe how we established our catalog of TriAnd M giant members; in \S\ref{spec}, the spectroscopic and kinematical properties of the expanded TriAnd M giant survey are presented; and \S\ref{disc} summarizes the results and offers one model to explain them.

\section{\label{phot}Photometric Properties of the Features}

\subsection{M Giant Sequences Apparent in Color and Magnitude}

To examine what substructures are apparent in TriAnd using the 2MASS catalog, we selected stars in the region $100^{\circ}<l<160^{\circ}$, $-50^{\circ}<b<-15^{\circ}$. Colors were restricted to $(J-H)_{0}>0.561(J-K_{S})_{0}+0.22$ and $(J-H)_{0}<0.561(J-K_{S})_{0}+0.36$, which should isolate a relatively pure sample of M giants \citep{maj03}. We targeted stars with $9.5<K_{S,0}<12.5$. A reddening restriction of $E(B-V)<0.555$ was applied to minimize the contribution of highly reddened sources close to the Galactic plane. We used the same methodology as \citet{maj03} to deredden the stars (i.e., an interpolated value of the \citet{schlegel98} maps applied to each star).    

\begin{figure}
\epsscale{0.7}
\plotone{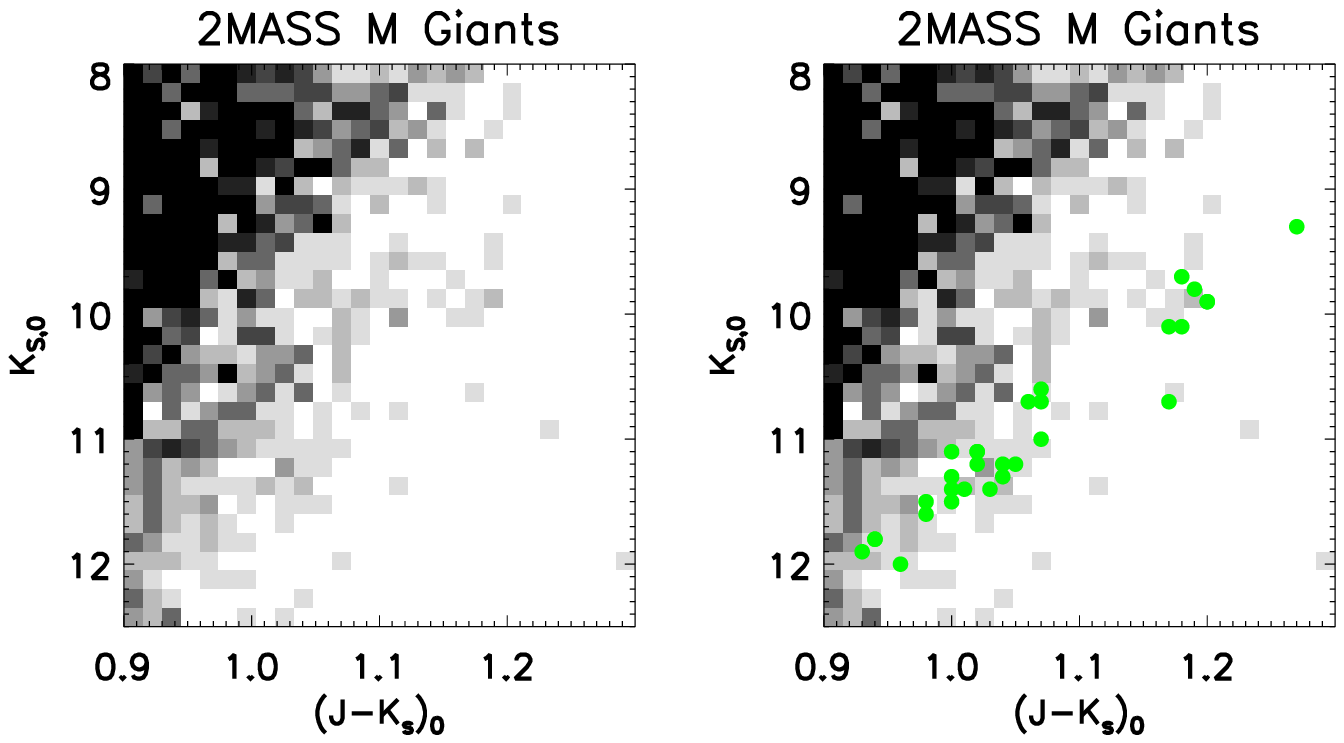}\\
\plotone{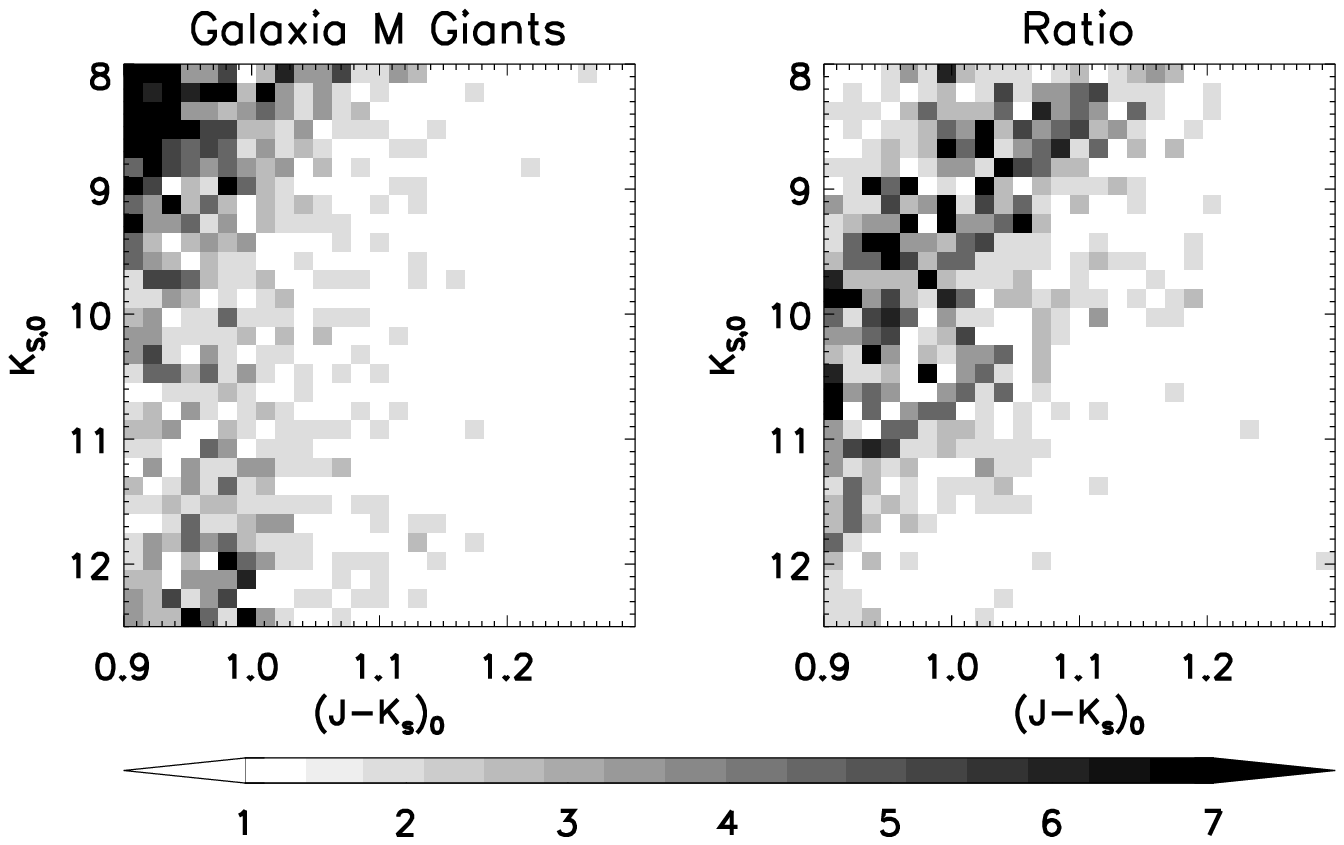}
\caption{\label{cmdmod}Hess diagrams of stars in the Triangulum-Andromeda region, $100^{\circ}<l<160^{\circ}$ and -$50^{\circ}<b<-15^{\circ}$. The upper-left panel shows 2MASS M giants, and the upper-right panel is the same but with the RP04 giants overplotted in green. The lower-left panel shows stars from a mock Galaxy generated by Galaxia, and the lower-right panel is the ratio of the two.}
\end{figure}

The upper-left panel of Figure \ref{cmdmod} shows a Hess diagram of the selected 2MASS stars with M giants from the RP04 study overplotted as green points in the upper-right panel.
For comparison, the lower-left panel shows the Hess diagram for this same region using output generated from the Galaxia synthetic model \citep{sharma11}, with identical $JHK_{S}$ color, magnitude, and reddening restrictions applied to both the observational and synthetic data.  
The bottom-right panel shows the ratio of the observed and synthetic data (2MASS/Galaxia). 
Three distinct RGB-like sequences are apparent from these comparisons: 
(i) the  brightest and densest sequence (emerging at  $(J-K_{S},K_{S})_0 \sim (0.9,9.5)$ and apparent in the bottom right panel) is GASS, which is known to extend into this part of the sky \citep{ibata03,crane03,rp03,martin07}; 
(ii) a second, slightly fainter sequence, seen emerging at $(J-K_{S},K_{S})_0$=(0.9,11.5) and spanning $10.5<K_{S,0}<11$ in the upper-left panel, traces out a clear RGB that stands out above the expected background level seen in Galaxia in the lower-right panel; 
and (iii) the locus of the RP04 giants (upper-right panel), that are known to be coherent in velocity, traces a third, fainter sequence, emerging at $(J-K_{S},K_{S})_0$=(0.9,12). 
%% AAS 28Apr
We note, in generating their spectroscopic sample from the apparent overdensity on the sky and in a range of apparent magnitudes, RP04 employed a probability density function (PDF) for M giants in 2MASS assuming that they followed a metallicity distribution centered at $-1.0$ dex with a dispersion of 0.4 dex. The distance PDF for a single star thus had a large spread due to this uncertain metallicity distribution. This large scatter in the distance estimates, combined with the aim of avoiding contamination from GASS and MW disk stars, caused RP04 to preferentially select the fainter RGB stars for follow-up spectroscopy.

\begin{figure}
\epsscale{1.0}
\plotone{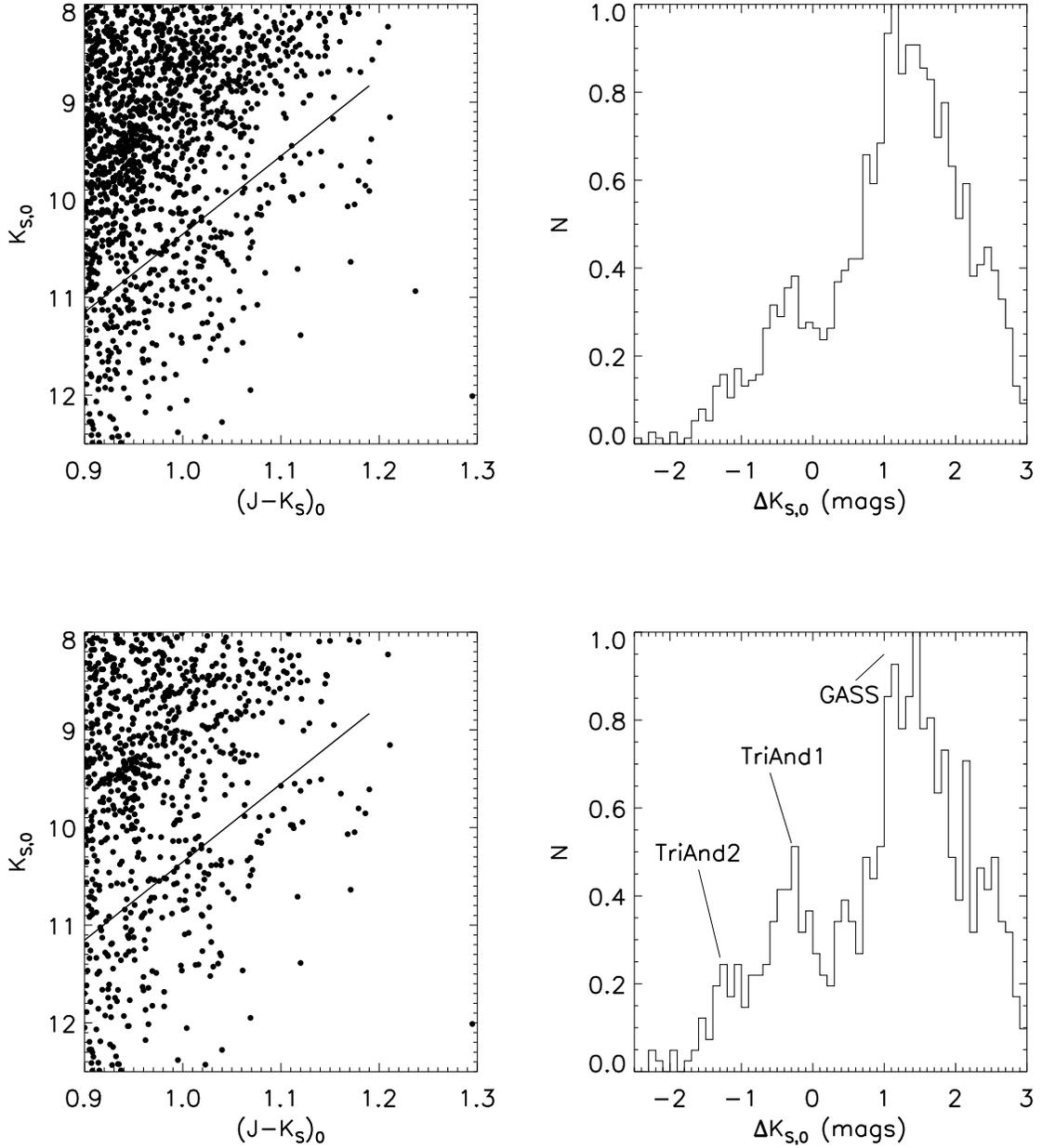}
\caption{\label{fiduc}Color-magnitude sequences in 2MASS. The left panels show 2MASS stars with $(J-H)$ and $(J-K_{S})$ cuts applied to isolate M giants; the solid line is a fiducial RGB selected to trace apparent overdensities seen in the 2MASS CMD (Figure \ref{cmdmod}). The distance between each star in the left panel and the fiducial RGB was computed, and the right panels show the histogram of these distances. In the top panels, the spatial region $100^{\circ}<l<160^{\circ}$, $-50^{\circ}<b<-15^{\circ}$ is shown, while the bottom panels show $100^{\circ}<l<160^{\circ}$, $-50^{\circ}<b<-20^{\circ}$; the slight change on the upper limit in $b$ is meant to reduce contamination from disk stars. Three peaks are seen in the right panels, from right to left corresponding to GASS, TriAnd1, and TriAnd2.}
\end{figure}

To assess whether the latter two (non-GASS) sequences can be more clearly distinguished, the magnitude difference, $\Delta K_{S,0}$, for each M giant from a linear fiducial RGB locus line is computed (see the left panels of Figure \ref{fiduc}); the fiducial line used has a slope that is aligned with the bright (non-GASS) feature emerging at $(J-K_{S},K_{S})_0$=(0.9,11.5) in the upper-left panel of Figure \ref{cmdmod}. The CMD in the top panel of Figure \ref{fiduc} shows stars selected from the same region as Figure \ref{cmdmod}, while the CMD in the bottom panel has a lower upper limit in Galactic latitude $b$ of $-20^{\circ}$. (There is expected to be significant contamination from disk stars between $-20^{\circ}<b<-15^{\circ}$, and the bottom panels improve the clarity of the features.) The right panels show histograms of the magnitude differences from the fiducial line. Overdensities corresponding to GASS (at $\Delta K_{S,0} \sim$ 1.5 mags) as well as the two fainter sequences (at $\Delta K_{S,0} \sim$ $-0.4$ mags and $\Delta K_{S,0} \sim$ $-1.2$ mags) are seen.

Overall, we conclude that there is an additional M giant population in the TriAnd region apparent in 2MASS that is clearly distinct from either the GASS or TriAnd sequences that have previsouly been identified in RP04.

\subsection{\label{iso}Isochrone Fitting} 
One obvious interpretation of the two RGB sequences seen in the 2MASS CMD in M giants in the TriAnd region is that they are the counterparts of the two MS features detected in \citet{martin07} and dubbed TriAnd1 and TriAnd2. 

\begin{figure}
\plottwo{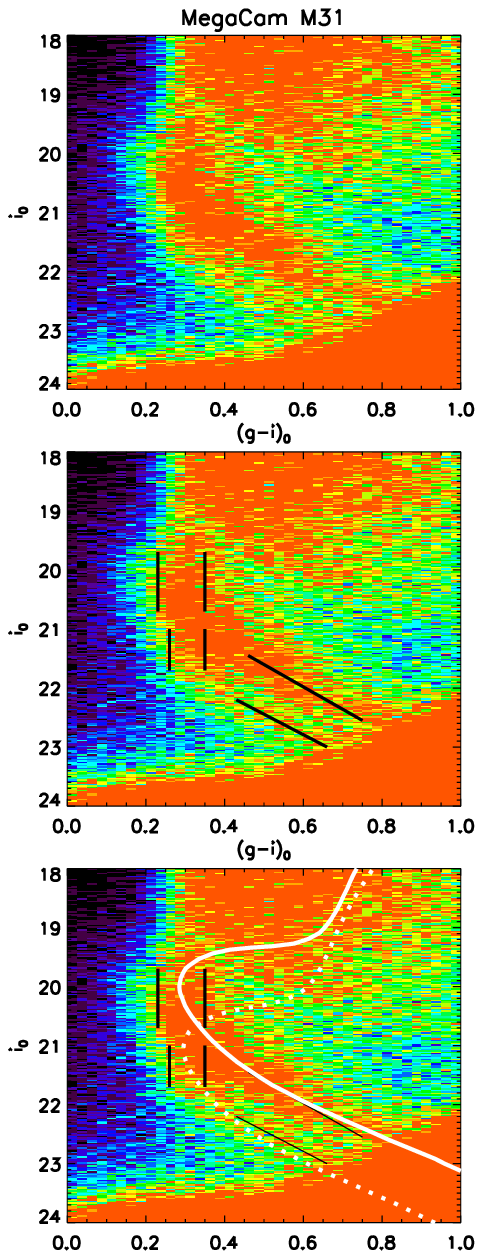}{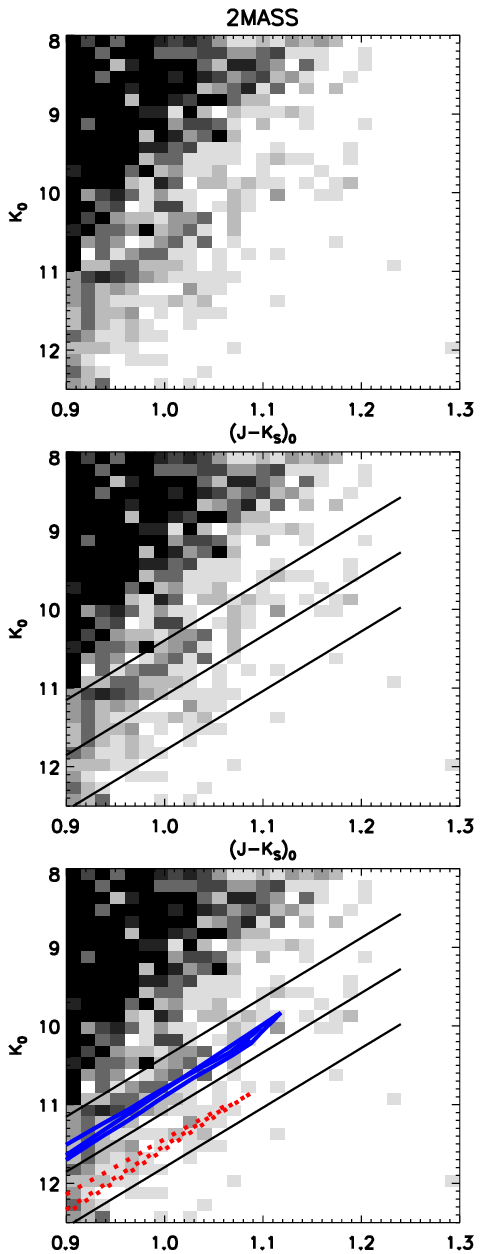}
\caption{\label{megaiso}Examples of isochrones that simultaneously fit the MegaCam $(g-i,i)_0$ main sequence data (left panels) and 2MASS red giant branch data (right panels). The isochrone shown for TriAnd1 is an 8 Gyr population with [Fe/H]=-0.8 at 18.2 kpc; that for TriAnd2 represents a 10 Gyr population with [Fe/H]=-1.0 at 27.5 kpc.}
\end{figure}

Under this assumption, we simultaneously fit isochrones \citep{bressan12} to the TriAnd1 and TriAnd2 MSs identified in the \citet{martin07} $(g-i,i)_0$ CMD\footnote{The $g_0$ and $i_0$ magnitudes were transformed from the MegaCam to SDSS system; the MegaCam data was kindly provided by Nicolas Martin.} and the M giant RGBs using isochrones in the SDSS and 2MASS $JHK_{S}$ photometric systems, respectively.  
The upper panels of Figure \ref{megaiso} show the MegaCam (left) and 2MASS (right) CMDs to which we fit the isochrones. The middle panels of Figure \ref{megaiso} show: on the left, the slanted black lines are the MS ridge lines to which we fit the isochrones and the parallel vertical black bars are the regions selected to constrain the color of the MS turn-off (MSTO) point; on the right, the loci isolate the regions used to constrain the 2MASS RGB/AGB fits around the apparent sequences. For a given isochrone, the RGB and AGB tracks differ by only $\sim$ 0.4 magnitudes for cool giants \citep[see, e.g., Figure 6 of][]{sheffield12}, so we cannot distinguish between these late evolutionary phases. The slope of the lines used to define the 2MASS TriAnd1/2 regions are the same as that of the fiducial in the left panels of Figure \ref{fiduc}.

To simultaneously fit the isochrones in the 2MASS and SDSS filters, we looked at a grid of isochrones with metallicities ranging from $-2.0$ to $0.0$ in steps of 0.1 dex. For each metallicity, we first found the distance modulus that matched to the MS ridge lines in the MegaCam CMD. Second, we found the age range that falls between the MSTO bars. Third, only isochrones that fit these two criteria and fell between the 2MASS RGB loci were kept.  

Only a small range of isochrones fit all three criteria, and Table \ref{tab1} lists these isochrones.  The range in possible solutions is fairly narrow for the simultaneous fit, in large part because of the requirement of both (the brighter and fainter) RGB sequences containing M giants: More metal-poor populations at ages spanning 5-12 Gyr have RGB tips that are bluer than $(J-K_{S})_0$=0.9. Isochrones that fall into the acceptable ranges reported in Table \ref{tab1} are shown in the bottom panels of Figure \ref{megaiso}, where the solid lines are for TriAnd1 and the dotted lines for TriAnd2. Magnitude spreads of $\Delta m \sim \pm 0.25$ dex about the MS ridgelines fit to the two features in the MegaCam CMDs are apparent. This suggests acceptable distance ranges of $\Delta d$ about the numbers in Table \ref{tab1}, where $\Delta d / d = (\Delta m \ln{10}) / 5 = 10\%-20\%$ (or $\Delta d \sim$ 2-4 kpc for TriAnd1 and $\Delta d \sim$ 3-6 kpc for TriAnd2).  
The estimates do indicate significant differences in distance (15-21 kpc versus 24-32 kpc) and age (6-10 versus 10-12 Gyr). The metallicity differences ([Fe/H] $\sim$ $-0.7$ to $-0.9$ versus $-0.9$ to $-1.1$) between TriAnd1 and TriAnd2 are less pronounced.  
All taken together, the age, distance, and [Fe/H] differences are suggestive of distinct populations, with the distance differences being the most robust. The derived distance of $\sim$ 28 kpc for the fainter MS/RGB sequences strongly suggests that these are in fact the same stellar population.  

\begin{figure}
\plotone{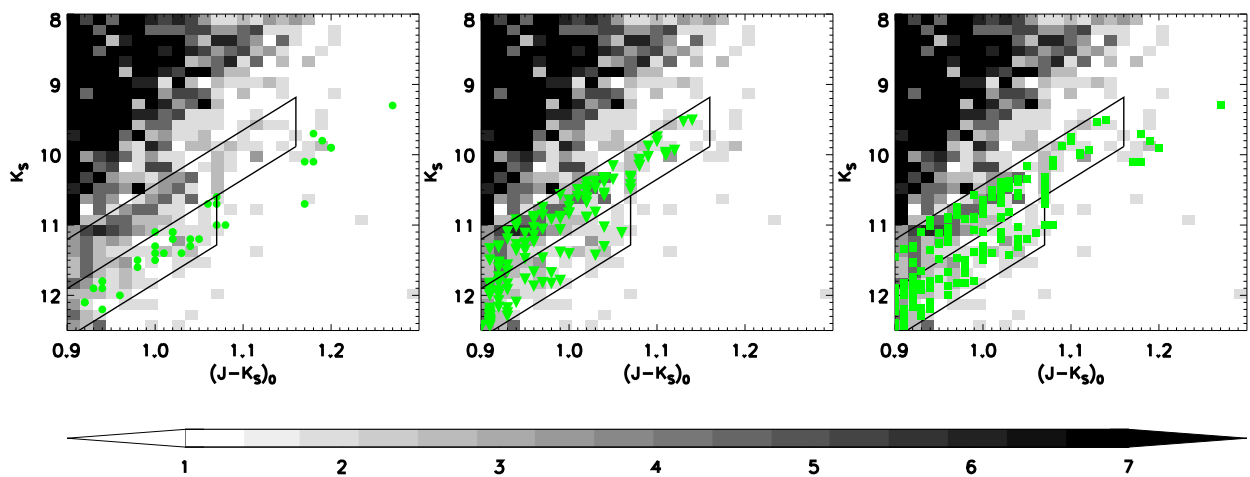}
\caption{\label{cmdobs}Hess diagram showing 2MASS stars with $(J-K_{S})_0>$ 0.9. The left panel overplots 2MASS TriAnd stars (both giants and dwarfs) from the RP04 study as green circles, the middle panel overplots the 170 2MASS stars explored further in this work as green triangles, and the right panel overplots the combined sample (RP04 stars and our expanded sample of 170 stars) as green squares. The boxes show the selection regions used to identidy members of TriAnd1 and TriAnd2, where the box extending to $(J-K_{S})_0$ of $\sim$ 1.16 corresponds to TriAnd1.}
\end{figure}

In Figure \ref{cmdobs}, we show the 2MASS CMD with $(J-K_{S},K_{S})_0$ boundary boxes for TriAnd1 and TriAnd2 overplotted. The $(J-K_{S})_0$ boundaries were informed by the results of the isochrone fitting: the TriAnd1 RBG ends at $(J-K_{S})_0 \sim 1.16$ and the TriAnd2 RGB ends at $(J-K_{S})_0 \sim 1.07$. On Figure \ref{cmdobs}, we also show the location of the stars targeted in this study and the RP04 giants on the $(J-K_{S},K_{S})_0$ CMD. 

\section{\label{data}Defining the Sample}

After establishing the presence of multiple, distinct RGB sequences in the TriAnd region, we now wish to explore those RGB features spectroscopically.
Figure \ref{cmdobs} summarizes our spectroscopic targets, where we show program stars with $(J-K_{S})_{0}>$ 0.9 (see Section \ref{mdwarfs} for details).
Targets were selected from 2MASS to fall around the sequences seen in Figure \ref{cmdmod}; overall, 170 stars were observed. 
When combined with the 36 stars (both dwarfs and giants) from RP04, we analyze 206 total stars in this study.       

\subsection{\label{red}Observations and Data Reduction}
Spectra for this work were collected over four observing runs, which are summarized in Table \ref{runs}. The November 2011 MDM run was beset by bad weather and electronic issues (random noise due to a faulty cable, manifested as spurious spikes, was superimposed on the spectra) and thus only a handful of stars from that run are of reliable quality. (We note here that the observing sample was restricted to $(J-K_{S})_0>$ 0.90 after we determined that nearly all of the stars with $0.86 < (J-K_{S})_0 < 0.90$ were classified as M dwarfs based on the strength of the NIR Na doublet; see Section \ref{mdwarfs}.) Pre-processing of the spectra for all runs were carried out using the IRAF $ccdproc$ task.  Variations in the bias level along the CCD chip were removed using the overscan strip on a frame-by-frame basis.  Biases were taken at the beginning and end of each night to verify that there were no significant drifts in the pedestal level.  For wavelength calibration, XeNeAr lamp frames were taken throughout the night at the same position as each target, to account for telescope flexure.  
To account properly for fringing in the quartz flat fields in the red, a low-order polynomial was fit to the median-combined flats to create a normalized flat (using the $response$ task) and the science frames were divided by the normalized flat.
The $apall$ and $identify$ tasks were used for 1-D extraction and pixel-to-wavelength calibration.  The dispersion solution was applied using the $dispcor$ task.  

The 1-D wavelength-calibrated spectra were cross-correlated against standard stars using the $fxcor$ task, after running $rvcorrect$ on all of the spectra to account for the Earth's motion with respect to the barycenter of the solar system.  The location of the night sky emission lines between 8400 \AA\ to 8500 \AA\ were checked for any systematic offsets during each night or individual offsets due to telescope flexure.  The overall level of stability for the KPNO 2.1-m was $\sim$ 5 km s$^{-1}$, with no systematic variation.  The level was higher for the Hiltner spectra (for the Nov 2011 and Oct 2012 runs), with the variations shifted by up to $\sim$ 10 km s$^{-1}$ in some cases (these shifts, when present, caused the night sky lines to systematically appear at slightly bluer wavelengths).  The heliocentric radial velocities for all targets are presented in Table \ref{tab3}, where the velocity is the mean of the velocities found from running $fxcor$. On a given night, anywhere from 3 to 8 standard stars were observed; if a standard did not correlate well with the other standards -- based on the derived radial velocity -- then it was not included in the mean calculation. The errors reported by $fxcor$ typically vary by only $\pm$ 1 km s$^{-1}$ for the cross-correlation results with the standards, meaning that they are not an accurate measure of the true errors, so we do not compute a weighted mean velocity. Modspec was set up to cover the spectral range 7900 \AA\ to 9200 \AA\ and Goldcam covered 7500 \AA\ to 9000 \AA; both Modspec and Goldcam had a spectral resolution of $\sim$ 4 \AA. In this spectral region, the Ca II triplet is the dominant feature and we used these lines to derive radial velocities. The errors on the radial velocities presented in Table \ref{tab3} are the mean of the differences in the velocities found from cross-correlating the star with the radial velocity standards observed on that night\footnote{The radial velocity standards had a similar spectral type as the program M giants.}. Twenty program stars were repeat observations; these results are listed in Table \ref{dups}. The average dispersion of the 20 repeat observations is $\pm$ 7.1 km s$^{-1}$. 

\subsection{\label{mdwarfs}M Dwarf Contamination}

Despite color cuts applied to avoid the problem, contamination from M dwarfs is a concern for cool giant stars with $K_{S,0}>$ 12.5 \citep{maj03,rp04}.  As noted in \citet{bochanski13}, the contamination rate rises to 66\% for stars with a $K$-band limit of 17.1 (UKIDSS $K$ filter\footnote{UKIDSS has a faint limit of $K\sim18.2$.}), even for a conservative blue limit of $(J-K_{S})>$ 1.02. 

To remove dwarfs from our sample, we checked the strength of the Na I doublet ($\lambda \lambda$ 8183, 8195), which is gravity-sensitive and can be used to discriminate dwarfs and giants \citep[see][]{schiavon97}.  This was done by measuring the equivalent width (EW) of the Na I doublet in two ways: (i) by fitting a Gaussian to each line of the doublet simultaneously, and (ii) by numerical integration (G. Damke et al., in preparation). Because the doublet may be contaminated by a water-vapor telluric band (which peaks at 8227 \AA), we first applied a telluric correction to our spectra using the spectrum of the white dwarf star Feige 11 as a telluric template, in the IRAF task $telluric$. Afterwards, the spectra are normalized using the IRAF task $continuum$. This allows us to define the continuum as 1.0 for both EW measurement methods.  In method (i), the wavelength ranges covered by the Gaussians (to measure the absorbed flux from each line in the doublet) are 8172-8187 \AA\ and 8190-8197 \AA. Then, the fitted relation is used to measure the EW in the wavelength range 8172-8197 \AA. In method (ii), we used numerical integration to measure the EW in the bandpass 8179-8199 \AA.
 
\begin{figure}
\plotone{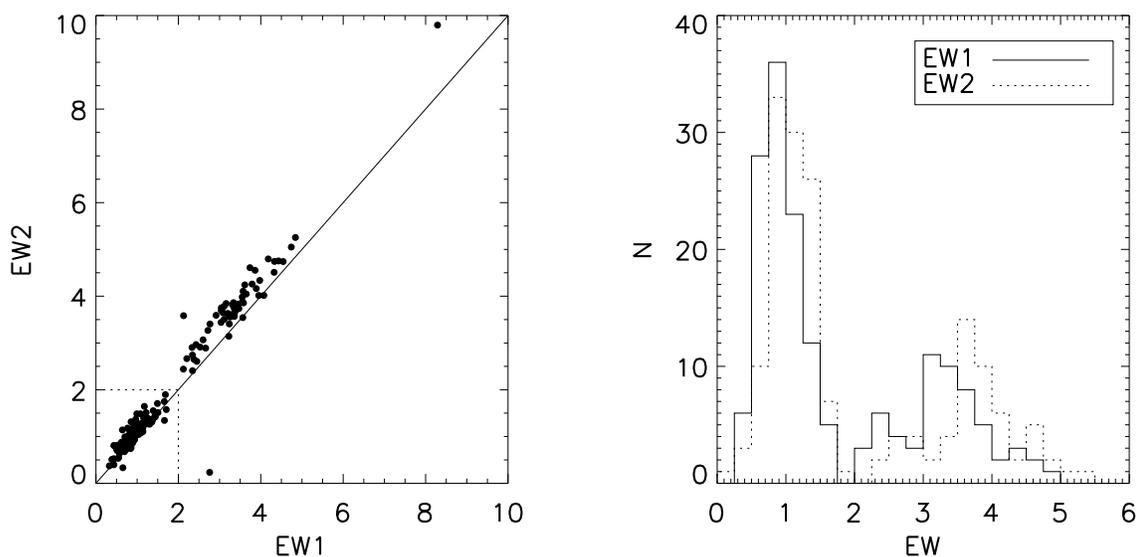}
\caption{\label{ew}A comparison of the two different methods for measuring Na I doublet equivalent widths (EWs), where EW1 is derived by simultaneously fitting two Gaussians to the doublet and EW2 uses numerical integration. The left panel compares the EWs derived using both methods, and shows that the two methods are highly correlated and more or less agree for almost all cases (the solid line is one to one, and the dotted lines show the apparent separation between dwarfs and giants). The right panel shows the distributions of EW1 and EW2. Two distinct populations are seen: stars with EW1 or EW2 less than $\sim$ 2.0 \AA\ are very likely giants. Figure \ref{rpmd} shows the reduced proper motion diagram for the program stars with UCAC4 proper motions color-coded by the strength of EW1.}
\end{figure}

Figure \ref{ew} compares these two equivalent widths individually (in the left-hand panel) as well as their distribution across the sample (in the right-hand panel). There is generally good agreement between EWs derived each way, although  there is a slight systematic shift of $\sim +0.3$ \AA\ for the EWs determined using numerical integration for stars with EW $\lesssim$ 2.0 (giants), and a bit higher shift (to 0.8 \AA) for the dwarfs. 
The amplitude of the Gaussian derived using method (i) is always fit to a positive value, thus avoiding an unrealistic fit to an emission line; this could explain the systematically lower values for the EWs found using the Gaussian fitting method.
Nevertheless, the methods generally agree and either can be used to discriminate the giants. 
The similarity of the estimates and the overall distributions suggests a cut requiring EW $<$ 2.0 to isolate giants.

\begin{figure}
\plotone{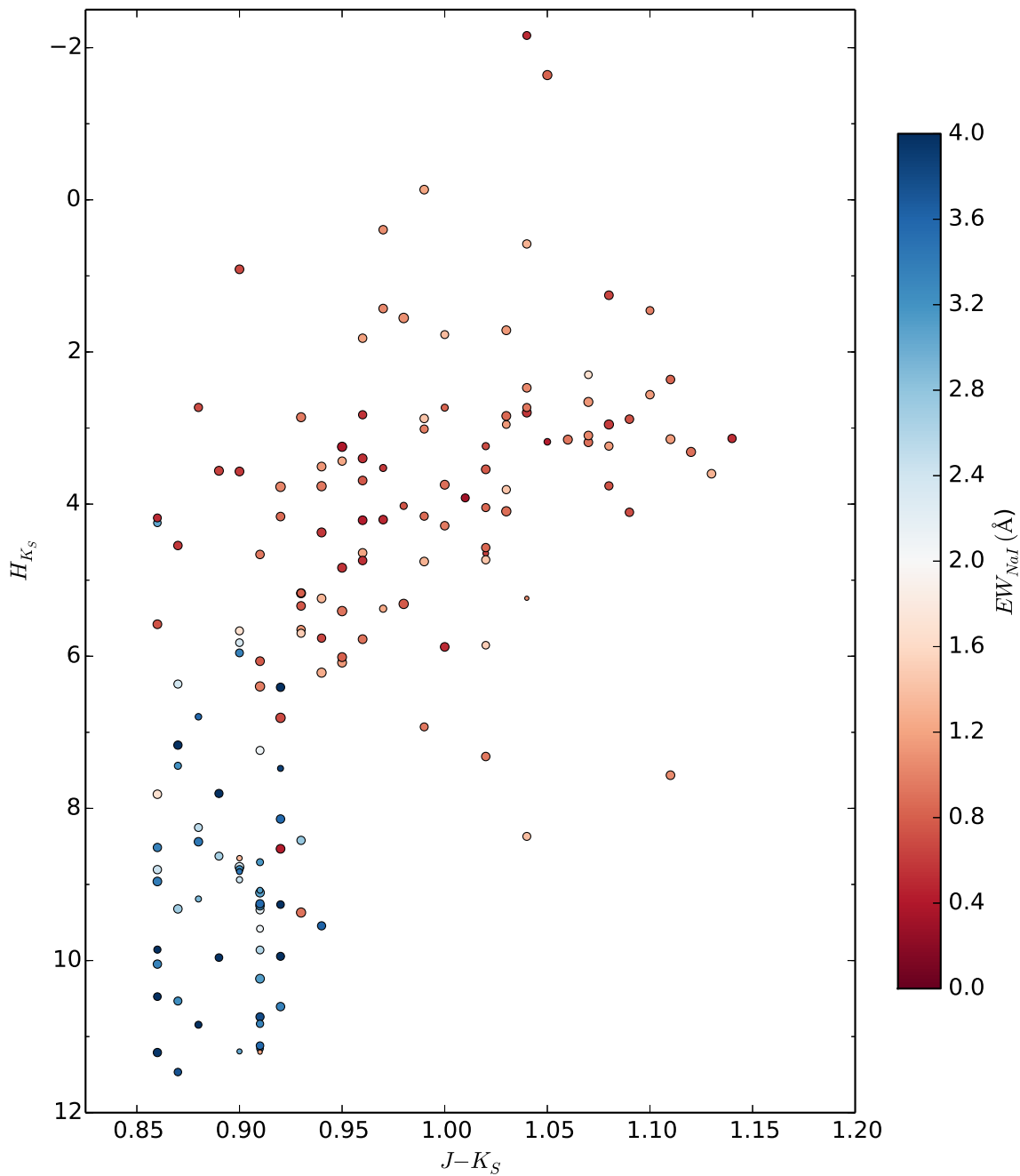}
\caption{\label{rpmd}The reduced proper motion diagram for the 158 program stars with UCAC4 proper motions. The points are color-coded according to the strength of the Na I IR doublet and the size of the point is inversely proportional to the error of $H_{K_{S}}$. Nearly all stars with an Na I doublet equivalent width strength greater than 2.0 fall into the dwarf star region of the RPMD (i.e., $H_{K_{S}}>6$).}
\end{figure}

To verify that the Na I doublet is a good discriminant of dwarfs and giants for our sample, and test the EW level chosen to make this distinction, we also looked at the location of the stars on the reduced proper motion diagram (RPMD), where $H_{K}=K_{S}+5\log{\mu}+5$. For $(J-K_{S}) \gtrsim$ 0.7, dwarfs can be separated from giants using the RPMD \citep[see, e.g.,][]{girard06}. We used proper motions from the UCAC4 Catalog \citep{zacharias13} to construct a RPMD to assess the correlation between the Na I doublet strength and luminosity class. Although the uncertainties on the proper motions for stars at these distances are quite large (typically 4 mas yr$^{-1}$, with a signal of 1-2 mas yr$^{-1}$ for giants), local dwarfs have proper motions at least an order of magnitude larger and thus can be reliably identified in a RPMD. Figure \ref{rpmd} shows the RPMD for the 158 program stars with UCAC4 proper motions available; the points are color-coded by the EW (found using method i) of the Na I doublet. These EWs correlate remarkably well with position in the RPMD: for EWs $\lesssim$ 2.0 (shown as the red points), stars overwhelmingly fall into the region populated by giants ($H_{K}<6$). 

To classify our sample stars, we tagged stars with either EW1 or EW2 less than 2.0 as a giant; 58 of the 170 targets were classified as dwarfs using this methodology.  
Our final catalog contains 142 M giants, with 30 giants from the RP04 sample and 112 newly identified TriAnd giant members (hereafter the ``S14'' sample).

\section{\label{spec}Spectroscopic Properties}

Drawing on our photometric and spectroscopic analyses of stars in the TriAnd region, we now focus on the stars that (i) fall into the TriAnd1/2 regions defined in Figure \ref{cmdobs} and (ii) are spectroscopically classfied as giants. Application of the additional color-magnitude cuts (i.e., the color boundaries for the TriAnd1 and TriAnd2 boxes shown in Figure \ref{cmdobs}) reduces the 142 giants identified in the previous section to 134. The fact that we have restricted our sample to colors of $(J-K_{S})_{0}>0.9$ means that our sample is inherently biased toward more metal-rich stars. Eight of the very red RP04 giants do not fall into the TriAnd boxes. However, these stars are very metal-poor (RP04 reports 4 of the 8 as having [Fe/H] $<-1.5$, and 3 of the remainder have [Fe/H] $\le -1.0$), suggesting that these stars are most likely carbon stars or in the thermally-pulsating AGB phase. 
In this section, we further refine our sample by applying an iterative clipping to the radial velocities of the giants falling within the TriAnd1/2 boxes to remove non-members of the TriAnd groups (Section \ref{rv}). Next, we estimate [Fe/H] for sub-samples of those giants with sufficiently high S/N spectra (Section \ref{feh}). Last, we use proper motions along with the estimated distances to TriAnd1 and TriAnd2 (from the isochrone fits) to estimate their transverse motions (Section \ref{pm}).

\subsection{\label{rv}Radial Velocities}

\begin{figure}
\epsscale{0.7}
\plotone{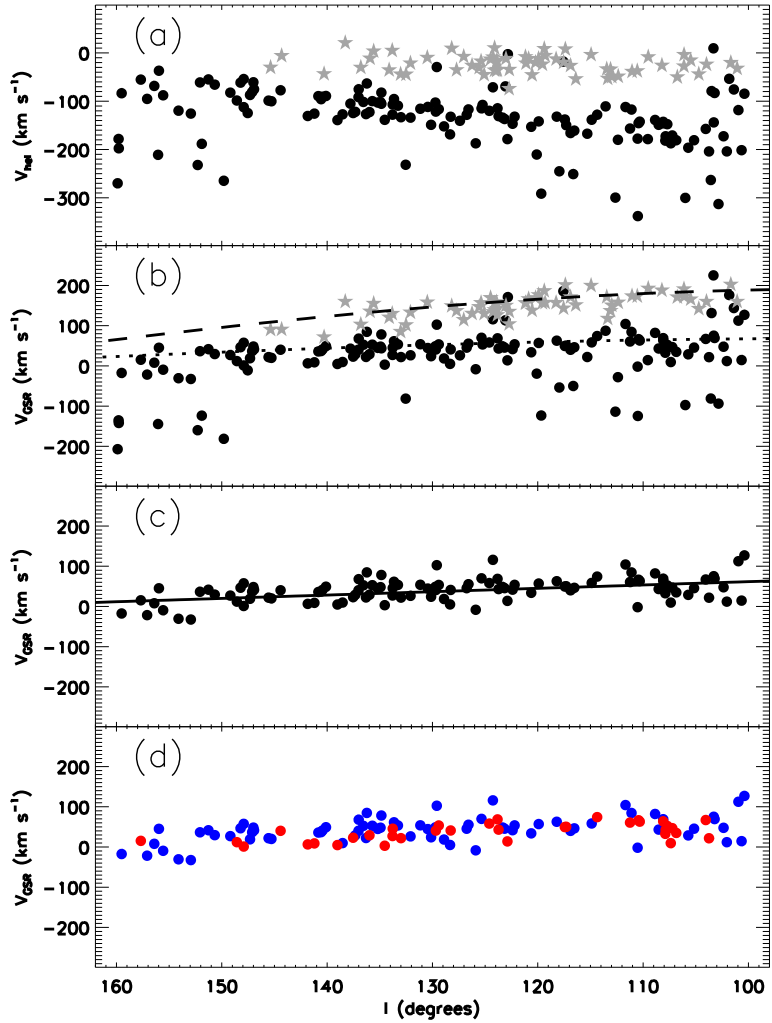}
\caption{\label{v4pan}The radial velocity distributions as a function of Galactic longitude for all stars observed as part of this program; all TriAnd stars from \citet{rp04} are also included in all three panels. In the top and middle panels, all program stars are shown; stars classified as dwarfs from the Na I doublet (see Fig. \ref{rpmd}) are shown as grey stars and those classified as giants are shown as black circles. The dashed black line (falling along the dwarf sequence) shows the circular orbit for local stars in the direction of TriAnd moving with $\Theta_0$=236 km s$^{-1}$, while the dotted line (falling along the giant sequence) is this same orbit for stars at $R_{0}$=25 kpc. In the bottom panels, a 2.5-$\sigma$ iterative clipping was applied to the giants and the ``cleaned'' sample of 109 TriAnd giants is shown. The solid line in panel (c) is the derived polynomial fit used to reject outliers. The stars shown in blue in panel (d) are classified as members of TriAnd1 and those shown in red are classified as members of TriAnd2.}
\end{figure}

RP04 found a trend in the radial velocties of the M giants observed in the vicinity of the TriAnd density peaks (see their Figure 4). For our extended spectroscopic survey, the M giants follow this same trend, as shown in Figure \ref{v4pan}. In panel \ref{v4pan}(a), the distribution of heliocentric radial velocities for all program stars (including the stars from the RP04 study) is shown, with stars classified as dwarfs in grey and those classified as giants in black. Panel \ref{v4pan}(b) shows the radial velocities but now in the GSR frame\footnote{We adopted the values $\Theta_{0}$=236 km s$^{-1}$ \citep{bovy} and ($U_\odot$,$V_\odot$,$W_\odot$)=(11.1,12.24,7.25) km s$^{-1}$ \citep{schonrich} to correct for solar motion.} (at rest with respect to the Galactic Center, to account for the total motion of the Sun); the dashed line plotted over the dwarf stars shows the expected trend for stars moving locally at $\Theta_{0}$=236 km s$^{-1}$, and the dotted line represents a circular orbit for stars at 25 kpc. In panel \ref{v4pan}(c), the results of applying a 2.5-$\sigma$ iterative clipping to the stars identified as giants are shown. The iterative clipping was done by fitting a first-order polynomial to $v_{\rm GSR}$ as a function of Galactic longitude and then removing 2.5-$\sigma$ outliers iteratively. The iterative clipping leaves 109 stars classified as members of TriAnd based on their $v_{\rm GSR}$ trend in $l$ (18 RP04 giants and 91 S14 giants); the polynomial used to reject outliers is shown as the black solid line in Panel \ref{v4pan}(c). In panel \ref{v4pan}(d) the giants from panel (c) are color-coded by their membership in TriAnd1 (blue circles) or TriAnd2 (red circles). 

\begin{figure}
\epsscale{1.0}
\plotone{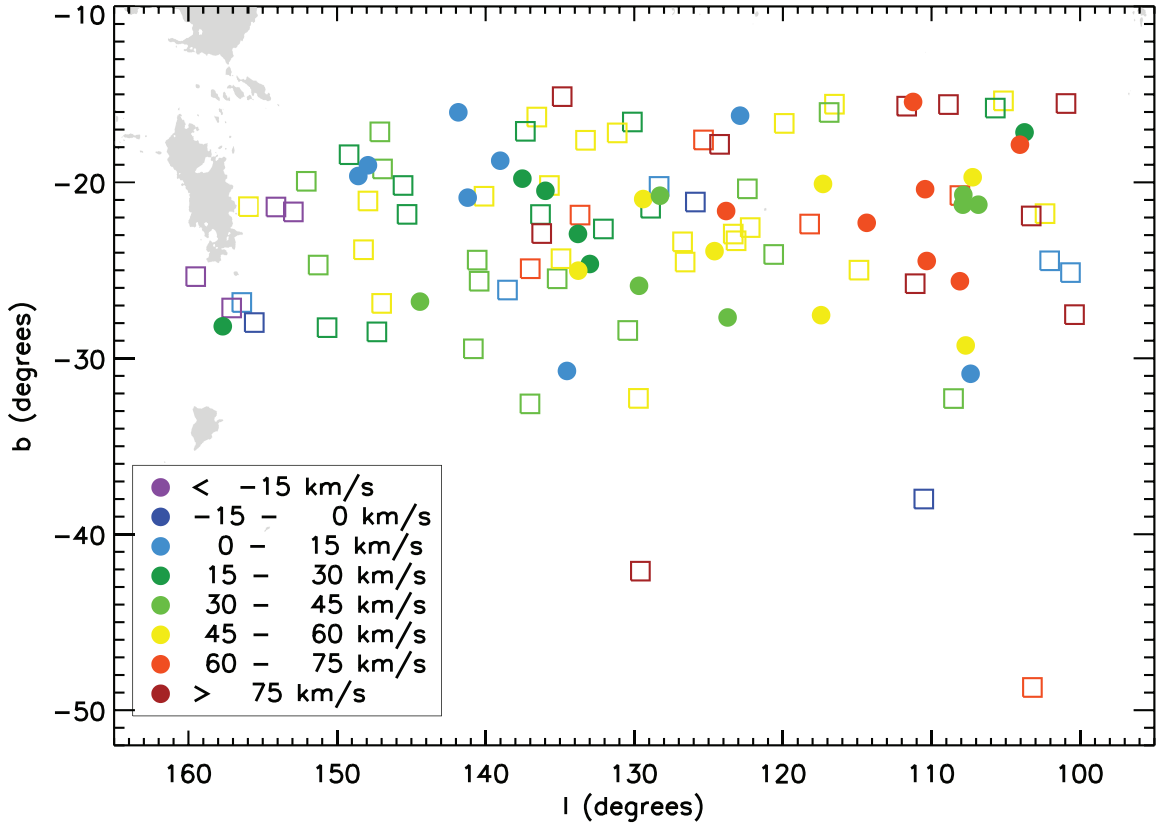}
\caption{\label{vgsrlb}The spatial distribution of the program M giants and those of \citet{rp04}, color-coded by $v_{\rm GSR}$, where squares represent stars photometrically classified as belonging to TriAnd1 and circles stars belonging to TriAnd2 (the color-magnitude boundaries for TriAnd1 and TriAnd2 are shown in Fig. \ref{cmdobs}). A gradient in $v_{\rm GSR}$ is seen as a function of $l$, and we also find the $b$ distribution to be more restricted (a lower $b$ limit of $-35^{\circ}$ is implied). As in Fig. \ref{subs}, the filled light grey regions correspond to areas on the sky with $E(B-V) > 0.555$.}
\end{figure}

Figure \ref{vgsrlb} shows the spatial distribution of the TriAnd M giants color-coded by their radial velocities (in the GSR frame).  
From Figure \ref{vgsrlb}, we see the velocity gradient in the sense that $v_{\rm GSR}$ is decreasing as the stars approach the Galactic Anti-Center (this is also seen in Figure \ref{v4pan}(c)).  We also note that almost all of the program stars observed with $b<-35^{\circ}$ (see Figure \ref{subs} for the spatial distribution of the targets) have been eliminated: of the 31 program stars with $b<-35^{\circ}$, only three remain after removing dwarfs (20 removed) and radial velocity outliers (8 removed). This suggests a lower spatial limit in Galactic latitude for the TriAnd overdensity.

\begin{figure}
\plotone{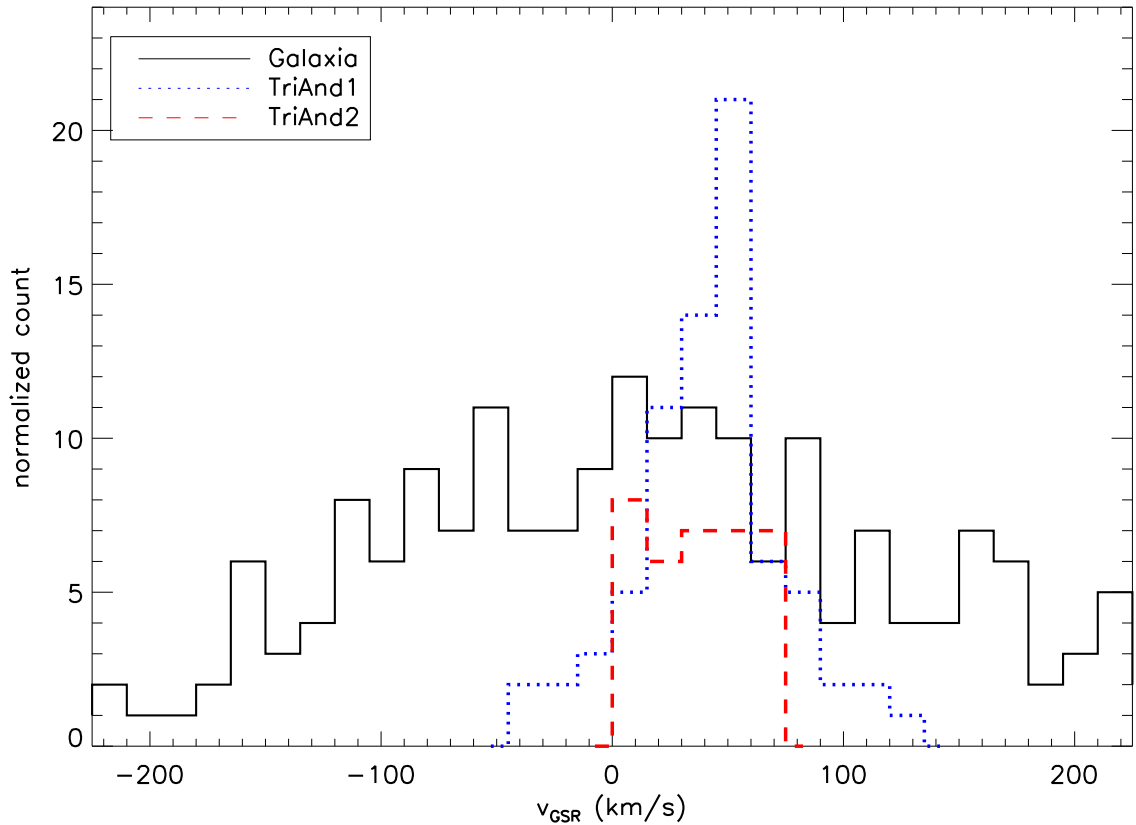}
\caption{\label{modrvs}The distribution of $v_{\rm GSR}$ for the stars shown in the bottom panel of Fig. \ref{v4pan}, with TriAnd1 giants plotted as the blue dotted line and TriAnd2 giants plotted as the red dashed line. The distribution of $v_{\rm GSR}$ for stars in a mock galaxy generated by Galaxia -- with identical photometric and spatial properties as the observed M giants -- that fall into the $(J-K_S,K_S)_0$ TriAnd1 and TriAnd2 selection boxes (see Fig. \ref{cmdobs}) is shown as the black solid line.}
\end{figure}

As with the photometric identification of the features, we can compare the kinematical properties of the features to those expected for random field thick disk and halo stars in the Milky Way by looking at the $v_{\rm GSR}$ distribution for a synthetic sample of stars generated from the Galaxia model \citep{sharma11}.  Figure \ref{modrvs} shows the distributions of $v_{\rm GSR}$, where we separate the $v_{\rm GSR}$ distribution for the 109 M giant TriAnd members by whether they fall into the TriAnd1 or TriAnd2 boxes, with TriAnd1 shown as the blue dotted distribution and TriAnd2 as the red dashed distribution.  An apparent difference in the $v_{\rm GSR}$ distributions is seen: TriAnd2 has a median near 40 km s$^{-1}$ and a fairly cold dispersion of $\sigma \sim$ 25 km s$^{-1}$, whereas TriAnd1 stars show a prominent peak at $v_{\rm GSR}$=50-60 km s$^{-1}$ and a similar dispersion. These dispersions are colder than expected for a random distribution of halo stars, but hotter than the $v_{\rm GSR}$ dispersion for the Sgr tidal stream ($\sigma \sim$ 10 km s$^{-1}$; \citet{maj04}) and the Orphan stream ($\sigma \sim$ 10 km s$^{-1}$; \citet{newberg10}).

A two-sided KS-test of the TriAnd1 and TriAnd2 $v_{\rm GSR}$ distributions results in a p-value of 0.65, which means that we cannot reject the null hypothesis that the two distributions were drawn from the same population. We also checked the kurtosis for each distribution: the $v_{\rm GSR}$ distribution for TriAnd1 is leptokurtic, with a kurtosis of 0.52, while that for TriAnd2 is platykurtic, with a kurtosis of -1.29. The two-sided Anderson-Darling test was applied to the TriAnd1/2 $v_{\rm GSR}$ distributions, as this test is more sensitive to differences in the tails of the distributions \citep{feigelson12}; the p-value for the Anderson-Darling test is 0.30. Although the p-value from the Anderson-Darling test is lower, we still cannot reject the null hypothesis.

The black solid line in Figure \ref{modrvs} shows the $v_{\rm GSR}$ distribution for a mock Galaxy with the same $JHK_{S}$ color-color and coordinate filters as applied to the observed stars, but now further restricted to show stars that fall into the TriAnd1 and TriAnd2 CMD boxes (there are 206 stars in the subset of the mock galaxy after applying these restrictions).  It is apparent that the dispersion of $v_{\rm GSR}$ for stars in the mock distribution is much hotter than that for the observed stars. 

We conclude that while TriAnd1 and TriAnd2 sit at larger Galactocentric radius than known disk populations, they are much more dynamically cold than the expected random halo population predicted by Galaxia. 

\subsection{\label{feh}Metallicity and Distance Estimation}

As an approximate and independent check on the metallicities derived from photometry using isochrones in Section \ref{iso}, we used spectral indices to derive metallicities for a subset of the S14 sample with sufficient S/N to give reliable results. 

All spectra taken for this study include the near-IR Ca II triplet. Several studies have explored the relation between the Ca II triplet and [Fe/H] for stars using spectral indices \citep{diaz89,cenarro01,du12,cesetti13}. The Paschen series causes blending in the region of the spectrum between 8360 \AA\ to 9000 \AA; however, as shown in \citet[][see their Figure 1]{cenarro01}, this blending is most prominent for hot stars (spectral types A and F). For stars cooler than spectral type M4 \citep{cenarro01,cesetti13}, molecular contamination affects both the continuum level and the flux in the region of the Paschen series lines. The color range for our program stars is $0.90<(J-K_{S})_0<1.14$ (we note that $<(J-K_{S})_{0}>$ = 0.99 for the S14 giant sample, which corresponds to $(J-K)_{CIT/CTIO}=0.95$, a color that equates roughly to spectral type M1.5 \citep{houdashelt00}). Considering the relatively minimal effect of the Paschen lines on the Ca II triplet in the color range probed by our study, we chose to use a simple sum of the three Ca II spectral indices.  We tested two different methodologies, one that includes the Paschen lines \citep{cenarro01} and one that is a simple sum \citep{du12}, and we found that a simple sum gave the best results (as measured by the mean difference between the published and derived metallicities for eight metallicity calibrators).

To compute the spectral index around each of the Ca II triplet lines, we found the total intensity of light within three central bandpasses, one covering each Ca II line. The spectral indices are pseudo-EWs measured in \AA\ (``pseudo,'' because the resolution is not high enough to measure a true EW). For each line, two bands flanking the central bandpass were also measured, to appropriately account for the continuum locally. The central and continuum bandpasses used are those from \citet{du12}.  The EW in \AA\ for each line is defined as \begin{equation}EW= \int_{\lambda_{1}}^{\lambda_{2}}\left(1- \frac{F_{l\lambda}}{F_{C\lambda}} \right)d\lambda \end{equation} where $F_{l\lambda}$ is the total intensity of the line between $\lambda_{1}$ and ${\lambda_{2}}$ and $F_{C\lambda}$ is the continuum flux and is computed as the interpolation between the red and blue bandpass centers to the center of the central bandpass (using the IRAF task $sbands$).

\begin{figure}
\plotone{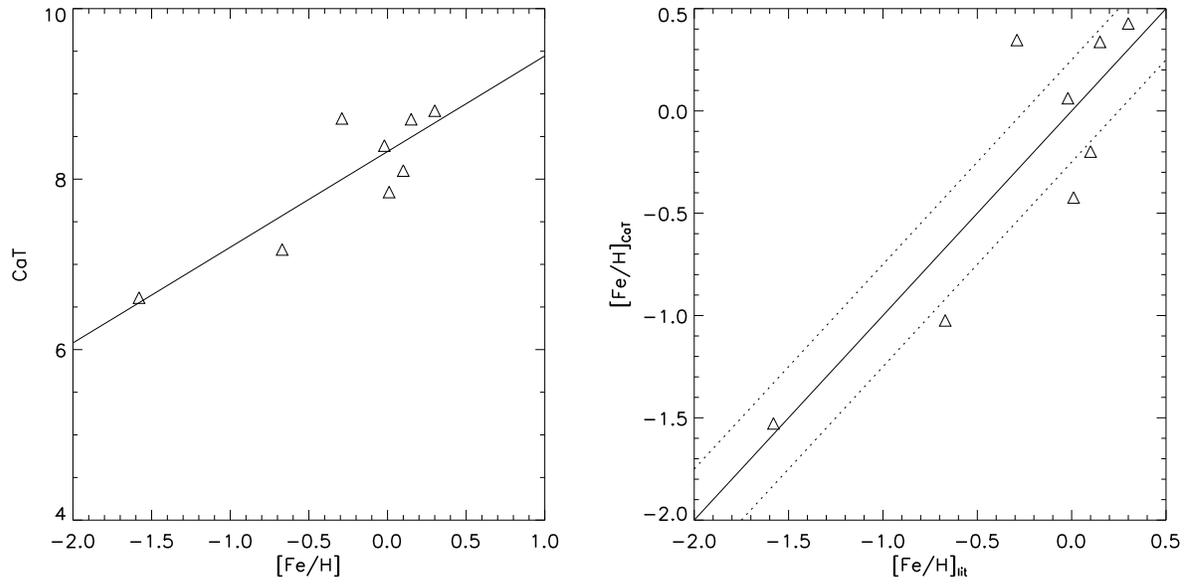}
\caption{\label{cat_std}The CaT-[Fe/H] linear fit for eight metallicity calibration giant stars, where CaT is the spectral index from \citet{du12} containing the near-IR calcium triplet. The dashed lines in the right panel are $\pm$ 0.25 dex away from the one to one (solid) line.}
\end{figure}

To test this approach for our own data, eight late-type giants with known [Fe/H] -- spanning the range $-1.7<$ [Fe/H] $<0.3$ and $0.92<(J-K_{S})<1.22$ -- were observed with Modspec on the Hiltner 2.4-m in June 2012 to define an empirical relationship between CaT (i.e., the sum of the Ca II triplet lines) and [Fe/H]. The [Fe/H] standards were taken from two sources: The PASTEL Catalog \citep{soubiran10} and the Astronomical Almanac. Our derived CaT-[Fe/H] relation for the 8 standard stars is shown as the solid line in the left panel of Figure \ref{cat_std}. In the right panel of Figure \ref{cat_std}, the [Fe/H] derived from the CaT lines is plotted against the published [Fe/H] values; the mean difference between the derived and published values of [Fe/H] is 0.27 dex. The estimated error in the derived metallicities is $\pm$ 0.30 dex, considering that most of the derived [Fe/H] values for the standards fall within 0.30 dex of the published values. 
For the standards observed on different nights of the June 2012 run, the variation in CaT is on the order of 0.05 \AA.

We next applied this method to program stars with S/N$>$20, hereafter the ``CaT stars'' (several stars with S/N$>$20 could not be used, because a cosmic ray fell within one of the spectral bandpasses needed to compute the EW). \citet{du12} show how the CaT-derived metallicities degrade with low S/N. Sixty-one CaT stars are classified as members of TriAnd1 ($<K_{S,0}>=10.7 \pm 0.57$) and 13 CaT stars are classified as members of TriAnd2 ($<K_{S,0}>=11.7 \pm 0.33$). 
The mean [Fe/H] derived from the CaT index for the 61 stars in TriAnd1 is $-0.62 \pm 0.44$ dex, where $\pm$0.44 dex is the standard deviation of the metallicities. 
The 13 CaT stars in TriAnd2 have a mean derived [Fe/H] of $-0.63 \pm 0.29$ dex. There is not a statistically significant difference between the spectroscopically derived [Fe/H] for TriAnd1 and TriAnd2 members. The metallicities for M giants classified as TriAnd1 members derived using the Ca II triplet lines agree with those derived from the isochrone fitting, within the estimated range of errors from the isochrone fits. 
Our value of [Fe/H]=$-0.62$ dex for TriAnd1 is similar to the value of $-0.64 \pm 0.19$ dex derived by \citet{chou11} in their high-resolution spectroscopic follow-up of 6 bright M giants from the RP04 study.  The M giants studied by \citet{chou11} have $JHK_{S}$ photometry that place them closer to TriAnd1 in the CMD. We have two giants in common with the \citet{chou11} and RP04 studies: 2333383+390924 and 2349054+405731 (both are listed in Table \ref{dups}). For 2333383+390924, we derive [Fe/H]=$-0.06$ (whereas \citet{chou11} derived [Fe/H]=$-0.63 \pm 0.11$ dex and RP04 derived $-0.1$ dex). 
For 2349054+405731 we derive [Fe/H]=$-0.22$ (whereas \citet{chou11} derived [Fe/H]=$-0.33 \pm 0.14$ dex and RP04 derived $-1.1$ dex). Our [Fe/H] value for 2333383+390924 agrees well with that derived by RP04 (but not very well with the value derived by \citet{chou11}), whereas we find good agreement with \citet{chou11} for 2349054+405731 but a large difference with the RP04 value. RP04 used a slightly different method to derive the spectral indices, so we may expect significant differences between our derived [Fe/H] and those from RP04. 

To derive distances, we used a refined version of the linear $M_{K_{S}}$-$(J-K_{S})$ relation derived for red giants presented in \citet{sharma10}, to account for metallicity dependency.  To do this, we found the best-fit line to 10 Gyr red giant isochrones (we note that the results are insensitive to the age) of [Fe/H]=0.0, $-0.5$, and $-1.0$ (M giants are an inherently metal-rich population and most will have [Fe/H] that fall within this range); as expected, a linear fit sufficed and was merely shifted up and down in $M_{K_{S}}$ to fit the RGBs of the different metallicity populations.  The relationship derived is 
\begin{equation}M_{K_{S}} = (3.8 + 1.3 {\rm[Fe/H]}) - 8.4 (J-K_{S}) \end{equation}

\begin{figure}
\plotone{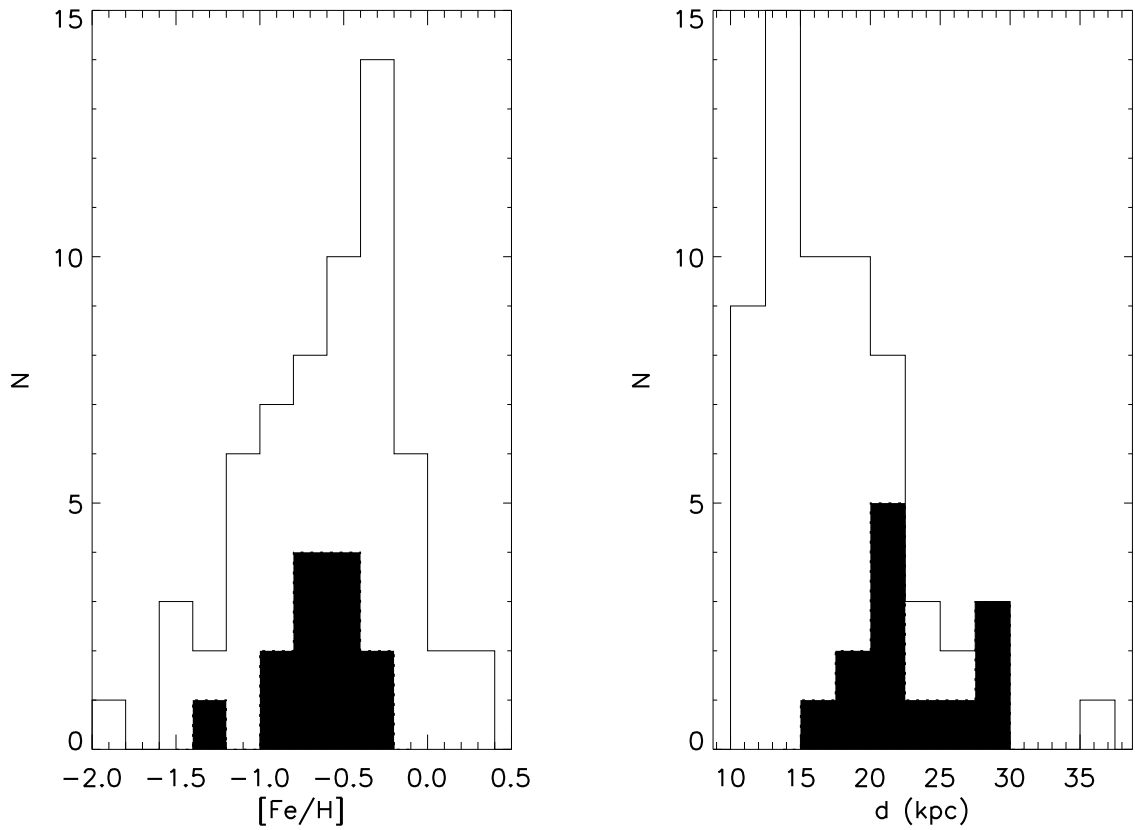}
\caption{\label{cat_d}The distributions of [Fe/H] and distance derived from summing the calcium triplet spectral lines. The open histograms contain stars classified as members of TriAnd1 and the filled histogram contains members of TriAnd2. We note that the distances are biased toward closer values, as we used sample stars with the highest S/N to derive the distances.}
\end{figure}

Distances were derived individually for each star using its color and estimated [Fe/H]. The mean distance and dispersion around the mean for the 61 CaT stars in TriAnd1 is 17.5 $\pm$ 5.1 kpc and for the 13 CaT stars in TriAnd2 is 22.6 $\pm$ 4.2 kpc; the distance distributions for TriAnd1 and TriAnd2 are shown in Figure \ref{cat_d}. The distances, particularly for TriAnd2, are biased toward closer stars since we are only using high S/N measurements and, thus, these are not an indicator of the true distance distribution. However, our distance estimates do support the finding (from the isochrone fitting) that stars in TriAnd1 are closer on the whole than those in TriAnd2.

\subsection{\label{pm}Proper Motions}
\begin{figure}
\plottwo{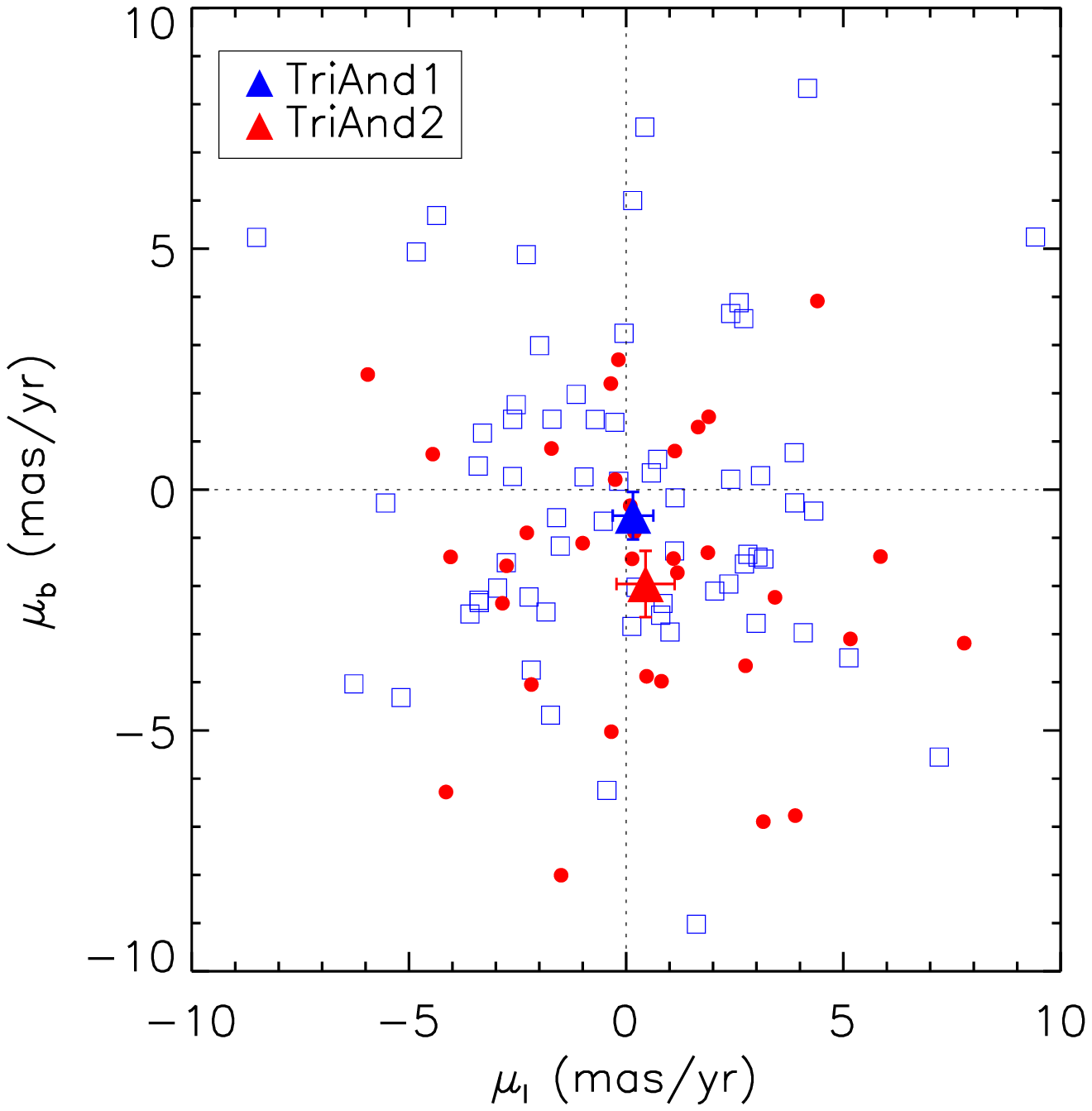}{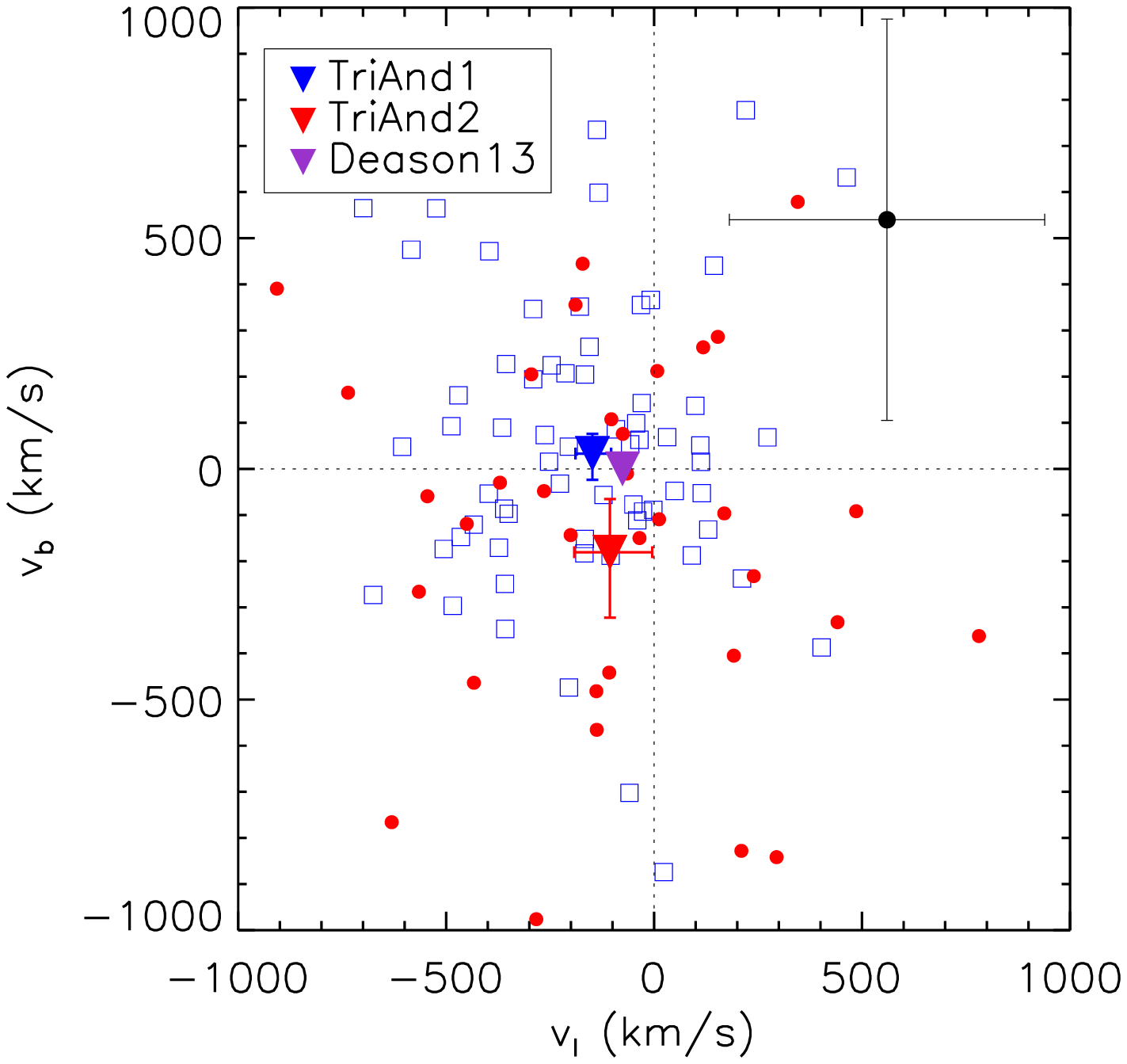}
\caption{\label{vlb}The left panel shows the distribution of $\mu_{l}$ and $\mu_{b}$ for observed M giant stars with UCAC4 proper motions in the TriAnd1 and TriAnd2 groups (blue open squares and red filled circles, respectively). The means are shown as the blue/red triangles. Seven stars that fall beyond the displayed range are not shown. The right panel shows the distribution of $v_{l}$ and $v_{b}$ for observed M giant stars in the TriAnd1 and TriAnd2 groups. The blue/red triangles show the centroids of $v_{l}$ and $v_{b}$ and the purple triangle is the halo group detected by \citet{deason13}. Six stars that fall beyond the displayed range are not shown. The black cross in the upper right corner of the left panel shows the mean errors on $v_{l}$ and $v_{b}$.}
\end{figure}

We matched the stars in this study to the UCAC4 Catalog \citep{zacharias13}.  The proper motions are not of high enough accuracy to derive individual space motions; however, taken in the aggregate, we can use these proper motions to assess any statistical differences in the tangential motions of stars in TriAnd1 and TriAnd2.  The left panel of Figure \ref{vlb} shows the distribution of $\mu_{l}$ versus $\mu_{b}$, with the error bars showing the error in the mean of the proper motions in each dimension for the 106 TriAnd1/2 members with UCAC4 proper motions available (we note that there are 7 stars with proper motions greater than $\pm$10 mas yr$^{-1}$ in one dimension falling outside the figure).  It is clear that we cannot distinguish between TriAnd1 and TriAnd2 based on their proper motions.

Using the proper motions, we also estimated the components of the tangential velocity, $v_{l}$ and $v_{b}$, in Galactic coordinates for M giants separated by their classification into the TriAnd1 and TriAnd2 groups.  To find the tangential velocities, first the projection of the Solar motion in the direction of each star was computed using the Solar motion components from \citet{schonrich} and the $\Theta_{0}$ value from \citet{bovy}.  The components $v_{l}$ and $v_{b}$ in the GSR frame were then found by vectorially adding the projected solar motion to each star and using the centroid of the distance range found from the isochrone fitting for the members of TriAnd1 and TriAnd2 ($d_{\rm TA1}$=18 kpc and $d_{\rm TA2}$=28 kpc):
\begin{equation}v_{b} = 4.74 \: d \: \mu_{b} + v_{b,\odot}  \end{equation}
\begin{equation}v_{l} = 4.74 \: d \: \mu_{l} \: \cos(b) + v_{l,\odot}  \end{equation}
The results, plotted in the right panel of Figure \ref{vlb}, shows a distribution that appears skewed towards $v_l < 0$, corresponding to prograde motion at these longitudes (note that the 6 of the 7 stars with proper motions greater than $\pm$10 mas yr$^{-1}$ also have tangential velocity components greater than $\pm$1000 km s$^{-1}$, showing that they are either dwarf stars nearby or have faulty proper motions).

Errors were combined in quadrature, using the errors in the proper motions from the UCAC4 Catalog and an estimated distance error of 25\% on the mean isochrone distances. The mean values of the errors in $v_{l}$ and $v_{b}$ are (379,435) km s$^{-1}$ and are shown as the black point with error bars in the upper right of the right panel of Figure \ref{vlb}; these large values show the statistical nature of this exercise.  

Because the distance estimates for individual M giants are very uncertain (and the errors are unlikely to be well-represented by a simple Gaussian), we did not estimate the mean tangential velocity from this distribution, but instead show the centroid components estimated from the average proper motions in the right panel of Figure \ref{vlb} combined with the middle of the distance ranges for TriAnd1 and TriAnd2 and the solar motion projected along the centroid of the TriAnd region, $(l,b)=(128^{\circ},-23^{\circ})$.
The errors bars on these centroid points indicate the effect of the range of possible distances found for these structures (15-21 and 24-32 kpc, respectively) when combined with the 1-$\sigma$ uncertainties indicated for the average proper motions in the right panel of Figure \ref{vlb}. The mean values found for $v_{l}$ and $v_{b}$ for the 13 halo stars studied by \citet{deason13} are shown as the purple inverted triangle in the left panel of Figure \ref{vlb}.

Overall, while the distribution of ($v_l,v_b$) for individual M giants is suggestive of the TriAnd structures being on prograde orbits, the distance and proper motion uncertainties are as yet too large for retrograde orbits to be excluded. Nor do we have clear evidence that the tangential motions of TriAnd1 is different from TriAnd2. From the error bars and the possible ranges, we cannot rule out an association with the \citet{deason13} halo group (this result is discussed further in \S\ref{shell}). 

\section{\label{disc}Summary and Interpretation}
Based on our expanded survey of M giants in the direction of the TriAnd stellar overdensity, we find that the stars have properties consistent with two distinct features that appear to be associated with the MSs detected by \citet{martin07} in their study of foreground dwarfs in the direction of M31. The first is at a distance of 15-21 kpc and the second at 24-32 kpc.  The isochrone fits (but not the spectroscopic CaT subsamples) suggest that the closer feature may be slightly more metal-rich.    
Despite these differences, TriAnd1 and TriAnd2 exhibit identical radial velocity distributions and trends with Galactic longitude. 
The distribution of tangential motions for individual M giants suggest they could be on prograde orbits about the Galactic center, 
but the proper motion measurements and distance estimates used to derive these velocities are sufficiently uncertain that retrograde orbits are not yet conclusively ruled out. 

\subsection{\label{others}Relation to Other Triangulum-Andromeda Detections}

\subsubsection{``The Triangulum Stream''}
\citet{bonaca12} detected a thin stream in the same region as our present study, which they refer to as ``The Triangulum Stream.''  Using a matched-filter isochrone fitting technique, they found the stream to be an old population (12 Gyr) at 26 kpc with [Fe/H]=$-1.0$.  These properties agree fairly well with those we derived for TriAnd2 and might suggest that the two features could be related.  Using spectroscopic data from the SDSS DR8, \citet{martin13} picked up the same stream as \citet{bonaca12} and rename it the Pisces Stellar Stream, based upon its location on the sky.  However, \citet{martin13} derive a spectroscopic metallicity of $-2.2$ for this thin stream and place it at a farther distance of 35 kpc; these stars have a kinematical signature of a $v_{\rm GSR}$ peak of 96 km s$^{-1}$.  The $v_{\rm GSR}$ of our features are significantly lower than this value so we do not believe the two to be associated.  However, there is a group of stars peaked at $v_{\rm GSR}$=50 km s$^{-1}$ in the SDSS plate analyzed by \citet{martin13} (see their Figure 4) that may be associated with our TriAnd features.  In the end, and including the vastly different spatial scales of the two structures, we can conclude that the Triangulum Stream is not associated with TriAnd1 or TriAnd2.  Furthermore, there are very few M giants in the 2MASS catalog in this region, which implies that the ``Triangulum Stream'' may indeed be metal-poor -- as found by \citet{martin13} -- and thus would not contain many (any) M giants.

\subsubsection{\label{shell}A shell of stars at 20-30 kpc?}
In a recent study of the velocity anisotropy of the Milky Way's halo, \citet{deason13} detected a group of 13 MS/TO Milky Way halo stars in the foreground of M31.  The stars have multi-epoch HST data and therefore extremely high-accuracy proper motions.  The heliocentric distances to the 13 Milky Way halo stars, which were found using weighted isochrone fitting, are all within 20-30 kpc, with a mean distance of 24 kpc; hence \citep{deason13} conclude that the stars are potentially part of a shell structure.  At $(l,b)=(121^{\circ},-21^{\circ})$, these stars may be also associated with the TriAnd1/2 features explored in this work.  From Figure \ref{vlb}, the estimated tangential motions of our TriAnd1 and TriAnd2 members overlap with the mean value for the 13 halo stars in the \citet{deason13} study; however, given the huge uncertainties, of course we cannot rule out this association based on $v_{l}$ and $v_{b}$. To make a more general assessment of the possibility of an association, we ask how many dwarf stars we would {\it expect} to fall in the \citet{deason13} sample given the number of M giants we detect in the region.

We estimate the expected density of M giants and MS/TO halo stars in the distance range covered by this study and the \citet{deason13} study by comparing the luminosity functions appropriate to the magnitude ranges spanned for each stellar population (i.e., $9.5<K_{S,0}<12.5$ for the giants and $21.8< m_{F814W} < 24.8$ for the dwarfs).  The spatial region studied by \citet{deason13} is much smaller than ours, so we must scale the numbers accordingly.  The field of view for the HST Wide Field Camera is $202 \arcsec \times 202 \arcsec$; for three pointings, this amounts to a total area surveyed of 0.0095 deg$^{2}$.  The total area we have surveyed is roughly 2000 deg$^{2}$.  The luminosity functions --- the absolute number of stars per unit mass for a Chabrier log-normal IMF --- in the HST/ACS WFC (F606W and F814W) and the 2MASS $JHK_{S}$ filters were taken from \citet{bressan12}. 
We assume a population with an age of 10 Gyr and [Fe/H]=$-0.8$ at a distance of 25 kpc (these are approximate averages for the two populations identified) to find the relative number of giants and dwarfs.
After appropriately scaling the number of stars per unit mass to account for the different areas probed, we find a ratio (N$_{\rm dwarfs}$/N$_{\rm giants}$)$_{\rm expected}$ of .0197, meaning there should be roughly 1 dwarf in the \citet{deason13} sample for every 51 giants in the region that fell in the 2MASS seletion boxes.  The estimated contamination rate for dwarfs for stars with $(J-K_S)_0>0.9$ (i.e., the randomly sampled region) is 40/175, or $\sim$ 23\%. 
The number of M giants in this region is 193$\times$0.77=149, where 193 is the number of 2MASS targets falling into the TriAnd1 and Triand2 boxes and we have scaled these numbers by the anticipated dwarf contamination rate.  Therefore, the observed ratio is 13/149 or roughly 1 dwarf for every 12 giants. Overall this suggests that while we cannot rule out that the \citet{deason13} sample could be part of the same dynamical substructure as TriAnd, the Deason stars are not part of the same stellar population and there is no conclusive evidence for association between them.

\subsubsection{\label{pandas}The PAndAS ``Field of Streams''}
The PAndAS photometric survey is an expansion of the CFHT/MegaCam survey of M31 \citep{martin07}, with coverage of $\sim$ 360 deg$^{2}$ on the sky. In their analysis of the PAndAS photometry, \citet{martin14} present density maps of four regions in the deep $(g-i,i)_0$ CMD covering the region extending roughly $110^{\circ}<l<135^{\circ}$ and $-35^{\circ}<b<-15^{\circ}$ in Galactic coordinates, a region that is entirely covered by our M giant survey. In the region at an estimated heliocentric distance of 17 kpc -- congruent with previous TriAnd1 detections -- \citet{martin14} find a structured area on the sky, with a thin stream (estimated physical width of 300-650 pc) cutting through the field from east to west, clearly extending beyond their coverage area in both directions. The authors call this feature the PAndAS MW stream and identify an overdensity of stars in the western portion of the stream as its possible progenitor. The MW stream has some properties that are consistent with the \citet{penarrubia05} simulations of the GASS feature. This, along with the thin extent of the stream, lead \citet{martin14} to suppose that the MW stream is due to the tidal disruption of a dwarf galaxy that was accreted on a low eccentricity, planar orbit. 

The PAndAS field at $D_{\Sun} \sim$ 17 kpc shows a higher background (``low-level substructure'') than the three other regions analyzed (see Figure 2 of \citet{martin14}) and it seems apparent that the TriAnd1 feature is present in this region. Similarly, the region at $D_{\Sun} \sim$ 27 kpc (Panel 4 in Figure 2 of \citet{martin14}) contains a wispy feature just south of M31 that appears to be one of the overdense regions in TriAnd2 (i.e., the overdensity in Figure 2 of RP04, centered roughly at $l\sim$ 115$^{\circ}$ and $b\sim$ -25$^{\circ}$).
Also noteworthy is the CMD of a field that overlaps the TriAnd2 region -- field (d) in Figure 4 of \citet{martin14}. Although the density map for this field looks rather sparse, a diffuse MS still emerges, suggesting that TriAnd2 covers a much larger extent on the sky than the MW stream (as expected by the density maps in M giants shown in RP04). 

Lastly, the heliocentric radial velocities (DEIMOS/Keck spectroscopy) for eight stars that overlap in the CMD with the MW Stream are clumped at $\sim -125$ km s$^{-1}$. This signal is consistent with the range of radial velocities that we find in the region we studied that overlaps the PAndAS field (see the black circles in the top panel of Figure \ref{v4pan}). This suggests that there may actually be an association between the MW Stream and TriAnd1 and/or TriAnd2. The recent work of Deason et al. (2014) further suggest a common origin between the TriAnd1/TriAnd2 features and the PAndAS MW stream.

\subsection{\label{sims}A Model for the TriAnd1 and TriAnd2 Sequences}

The discussion in the previous sections indicated that the current data are insufficient to rule out or confirm associations between the TriAnd sequences and other features identified in the region (for example GASS), so we do not think we are yet in a position to present a definitive model of all structures in the region.
Instead we restrict our attention to exploring the specific class of models in which the TriAnd sequences represent stellar debris from the disruption of a satellite galaxy. 
In particular, we show that a single satellite disruption is capable of simultaneously explaining  what we consider to be the most robust observed properties of TriAnd1 and TriAnd2 --- the {\it differences} in their distances as well as the {\it similarities} in their sky coverage and velocity trends.
(Other scenarios are discussed in \S \ref{limits}.)

Debris from satellite disruption is most dense, and most likely to be found at the apocentric positions, where the stars spend most of their time. 
For the same reason, this is also the orbital phase where distinct debris populations lost on separate orbital passages are most likely to appear coincident on the sky.
To investigate the plausibility of interpretating TriAnd1 and TriAnd2 as just such distinct debris populations from the same satellite, a series of simulations were run under the {\it assumptions} that: (i) the observed zero in line-of-sight velocities is indicative of debris sitting close to the orbital {\it apocenter} (rather than pericenter); and (ii) that the structure we observe represents debris leading the dead or dying satellite along its orbit.
(Note that the second assumption is somewhat arbitrary. However, if the structures were instead associated with trailing debris, then the leading counterparts would be mixed throughout the inner Galaxy and the lack of prior detections in the literature would need to be explained.)
Under these assumptions we can interpret
the observed gradient in velocities as a signature of stars flowing away from us and slowing down towards the orbital apocenter.
As a consequence,  both the morphology in RP04 (with an edge in density parallel to lines of constant Galactic latitude) and the sense of the gradient indicate that the debris should be moving predominantly in the direction of increasing Galactic longitude (i.e., $v_l >0$ in the GSR frame, along a retrograde orbit), so we further assume no motion of the debris in the middle of the field in the direction of Galactic latitude (i.e., $v_b=0$).

\begin{figure}
\plotone{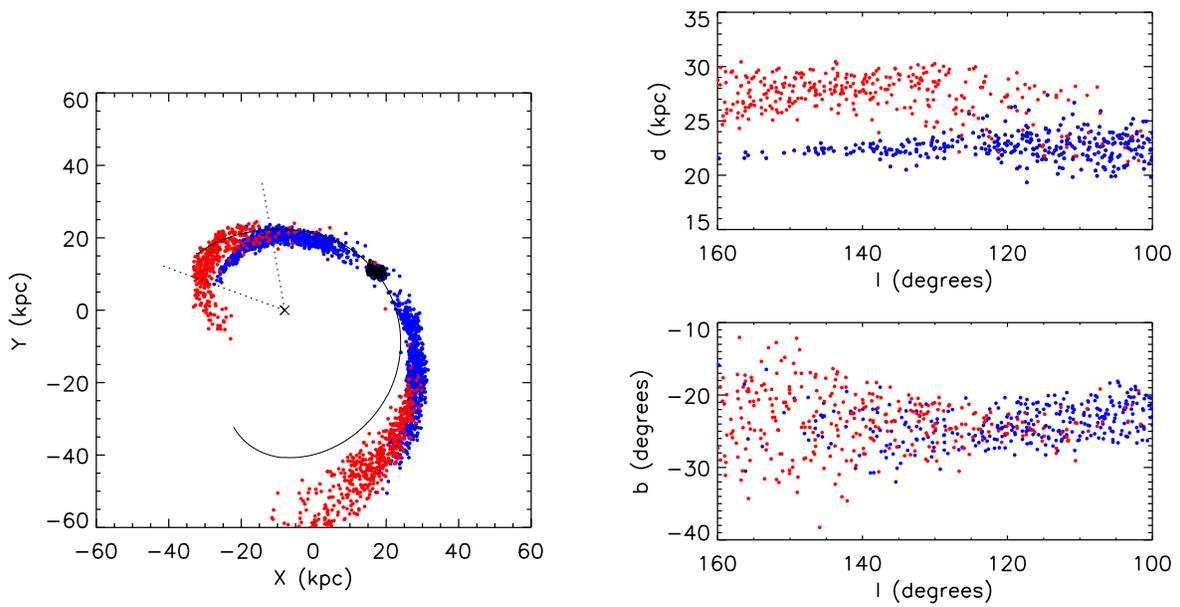}
\caption{\label{sims1}Results of an N-body simulation of a disrupting dwarf that can plausibly explain the TriAnd debris properties as we have measured them. `X' marks the position of the Sun and the dotted lines indicate the region observed in our survey. Blue points represent particles unbound from the satellite (shown in black) on the current pericentric passage and red points are the particles unbound on the previous passage.}
\end{figure}

Figure \ref{sims1} summarizes the results for one such simulation of the disruption of a $7 \times 10^8 M_\odot$ mass dwarf satellite system in orbit in a realistic Milky Way potential \citep{law10}. The dwarf in the simulation was represented by a Plummer model with scale-length 1.0 kpc, on an orbit of mild eccentricity with pericenters of $\sim 17$ kpc, apocenters $\sim 42$ kpc and radial time period of 0.7 Gyr. The model does not contain separate representations of the dark and light matter within the dwarf, so the positions of the simulated particles outline the extent of stellar debris expected in each projection once the extended dark matter halo is stripped to this mass and a significant fraction of the stellar component is being lost. In particular, the original total mass of the dark matter halo could have been significantly larger.

The left hand panel represents positions of particles projected onto the plane of the Galactic disk, lost on the previous pericentric passage of the satellite (blue) and the orbit beforehand (red). The cross indicates the position of the Sun and the dotted lines show the limits of the survey (i.e., $l=100^{\circ}-160^{\circ}$).

\subsubsection{Successes of the Model}

The lower right hand panel of Figure \ref{sims1} shows the positions of the red and blue particles projected onto the sky, demonstrating that the debris is in the appropriate location to represent TriAnd. The upper right panel shows that this single model can reproduce the bimodal distance distribution apparent in the data by appealing to the differences in position at apocenter for debris lost on different pericentric passages, with the closer, denser TriAnd1 corresponding to debris more recently lost. 

\begin{figure}
\epsscale{0.6}
\plotone{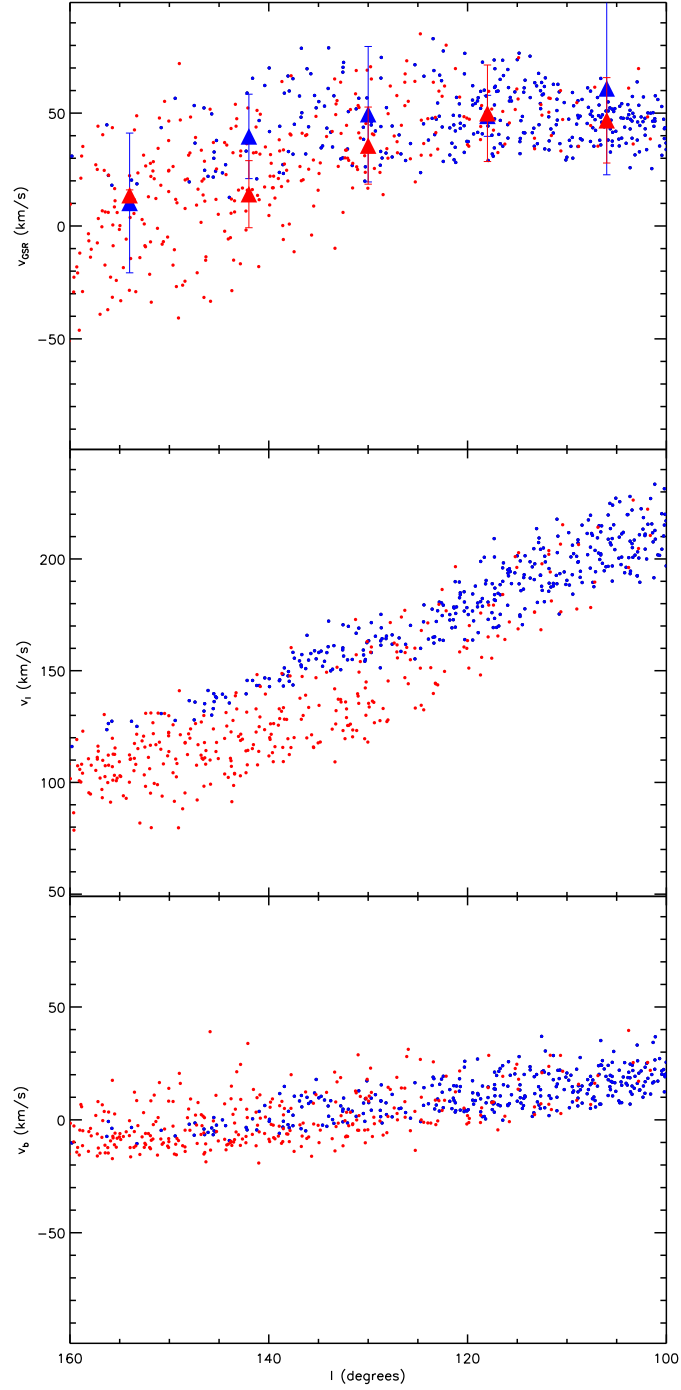}
\caption{\label{sims2}Mock observations of the simulations illustrated in Fig. \ref{sims1}. The triangles and error bars indicate the average and dispersion for Triand1 (blue points) and Triand2 (red points) in the real data.}
\epsscale{1.0}
\end{figure}

Figure \ref{sims2} summarizes the results in velocity-space. The upper panel shows that both red and blue simulated sequences (particles) follow the same observed trend in $v_{\rm GSR}$ (triangles and error bars representing the average and dispersion of the observed M giants). The lower panels reveal systematic differences in tangential velocities between TriAnd1 and TriAnd2, but these would be undetectable with the current accuracy of proper motions available.

Lastly, if the small abundance difference ($\sim$ 0.1 dex) between TriAnd1  and TriAnd2 suggested by the isochrone fits of Section \ref{iso} is confirmed, this can quite easily be explained in this picture as being due to a mild metallicity gradient in the progenitor object. 
The red particles (TriAnd2) occupied radii with an average and dispersion of 1.20$\pm$0.51 kpc within the model progenitor object used in the simulations, while the blue particles occupied radii with an average and dispersion of 0.95$\pm$0.42 kpc.
Thus the required metallicity difference would require a negative gradient of $\sim$ 0.5 dex kpc$^{-1}$. Gradients of this size have been seen in most sizeable dwarf spherical satellites of the Milky Way, including Sculptor and Leo II \citep[see Table 1 of ][]{kirby11}. Moreover, significant gradients in abundances have already been observed along the Sagittarius stream \citep{chou07,keller10}.

\subsubsection{\label{limits}Limitations of the Model and Alternative Explanations}

The model described above was selected after a modest exploration of parameter space with simulations where the tangential velocity in the region and total satellite mass were varied (while keeping the density, and hence fractional mass-loss-rate, constant).
Overall, it was found that: (i) the mass of the satellite at the point when observed debris is lost must be $\sim$ a few $10^8 M_\odot$ to reproduce the width of the velocity distribution  --- masses a factor of 5 higher or lower are inconsistent with the data; (ii) the tangential velocity of the debris must be $v_l \sim 75-125$ km s$^{-1}$ to reproduce the observed moderate velocity gradient. In particular, note that the morphology of debris in the left-hand panel of Figure \ref{sims1} is more stream-like (continuous density along the orbit) than cloud-like (specific concentration of debris at apocenter). The latter ``cloud-like'' morphology only appeared when a strong velocity gradient was produced in the simulations due to the satellite disrupting on a much more eccentric orbit.

While our model satisfactorily fits what we considered the most robust aspects of the current data (i.e. positions and line-of-sight velocities), it is far from a unique solution.
Moreover, it is mildly inconsistent with the (currently poor) estimates of tangential velocities whose distribution appears skewed towards negative $v_l$ (i.e., prograde orbits). 
Indeed, \citet{penarrubia05} found debris in tidal disruption models with similar velocity sequences moving in the opposite sense around the Galaxy.
In this case the zero in $v_{\rm GSR}$ in the debris occurs prior to the positive flow outwards along the orbit, so the debris 
would be at pericenter instead of apocenter. 
These {\it prograde} tidal disruption models are incapable of reproducing the distance offset between TriAnd1 and TriAnd2 during a single pericentric passage because the spatial distinction between debris lost on different passages is only apparent at apocenter.
TriAnd1 and TriAnd2 could instead correspond to different debris wraps on separate passges, but then the exact coincidence in $v_{\rm GSR}$ trends is hard to explain.
This scenario could be conclusively distinguished from our own with more accurate assessments of the proper motions of stars in the region.

A third possibility is to attempt to associate TriAnd1, TriAnd2  and  GASS all with the Galactic disk.
In particular, it has been previously pointed out that the TriAnd1 and TriAnd2 velocity gradient fits smoothly onto that observed for GASS at different Galactic longitudes (see RP04),  though GASS has different stellar populations and lies much closer to the Sun and the Galactic plane.
It would be interesting to explore whether all three features could simultaneously be produced by perturbing a self-gravitating disk system that includes a realistic population gradient. 
Such models would produce structures on prograde orbits.

\subsection{Conclusions}
We present an updated view of 2MASS M giants in the TriAnd region: We confirm additional members of the faint RGB sequence identified by \citet{rp04} and also identify a brighter RGB sequence, which we show to be likely associated with, but distinct spatially from, the faint RP04 sequence. The two distinct RGB features are directly compared with the two MS features detected by \citet{martin07} in the direction of Andromeda (TriAnd1 and TriAnd2). By simultaneously fitting isochrones to the 2MASS RGB features and the Megacam MS features, we estimate the age, distance, and [Fe/H] of each feature; we find significant differences between the age and distance of the two features -- the brighter, denser feature is younger and closer -- and a slight difference (on the order of 0.1 dex) in the metallicities of the features. The fainter MS detected by \citet{martin07} is consistent with being the optical counterpart of the RGB sequence detected by RP04.

Armed with our observed and derived properties of TriAnd1 and TriAnd2, we explore one possible origin scenario where the structures represent debris from the disruption of a satellite galaxy.
We find that a model with a progenitor satellite on a retrograde orbit that has been stripped over time to produce two distinct populations at the same orbital phase can explain the data. 
In these models, the  TriAnd feature is not morphologically a cloud but rather part of a more extended stream.
The observed gradient in $v_{\rm GSR}$ as a function of Galactic longitude is not steep enough to produce cloud-like morphologies \citep{johnston08}. Of course, it remains unclear if this is the only solution, and the association with and the nature of other structures along this line-of-sight \citep{penarrubia05,momany06} are still under scrutiny. 

\acknowledgements
This material is based upon work partially supported by the National Science Foundation under Grant Numbers AST-1312863, AST-1107373, and AST-1312196. A.A.S. thanks Nicolas Martin for sharing the MegaCam data, and Ting Li, Matthew Newby, David Hendel, Josh Peek, and Jeffrey Carlin for helpful conversations. A.A.S. and K.V.J. thank the anonymous referee for her/his feedback.
This work is based on observations obtained at the MDM Observatory, operated by Dartmouth College, Columbia University, Ohio State University, Ohio University, and the University of Michigan.

\clearpage

\begin{deluxetable}{c c c c}
\tabletypesize{\scriptsize}
\tablewidth{0pt}
\tablecaption{\label{tab1}TriAnd1/TriAnd2 Properties Derived from Isochrone Fitting}
\tablehead{
 & \colhead{[Fe/H]} & \colhead{$[d_{min},d_{max}]$} & \colhead{Age} \\
 & \colhead{dex} & \colhead{kpc} & \colhead{Gyr} \\
}
\startdata
TriAnd1 & -0.7 & 17-18 & 6-8 \\
	& -0.8 & 18 & 8 \\
	& -0.9 & 18-19 & 8-10 \\
TriAnd2	& -0.9 & 26 & 10 \\
	& -1.0 & 27 & 10-12 \\
	& -1.1 & 27-29 & 10-12 \\
\enddata
\end{deluxetable}

\begin{deluxetable}{c c c}
\tabletypesize{\scriptsize}
\tablewidth{0pt}
\tablecaption{\label{runs}Summary of Observing Runs}
\tablehead{
 \colhead{UT} & \colhead{Telescope} & \colhead{Spectrograph} \\
 }
\startdata
2011 Nov 11-12 & Hiltner 2.4-m & Modpsec \\
2011 Nov 15-20 & KPNO 2.1-m & Goldcam \\
2012 Oct 27-29 & Hiltner 2.4-m & Modspec \\
2013 Oct 19-20 & Hiltner 2.4-m & Modspec \\
\enddata
\end{deluxetable}

\begin{deluxetable}{c c c c c c c c c c c c c c}
\rotate
\tabletypesize{\scriptsize}
\tablewidth{0pt}
\tablecaption{\label{tab3}Properties of the Program Stars.}
\tablehead{
\colhead{ID} & \colhead{l} & \colhead{b} & \colhead{$K_{S,0}$} & \colhead{$(J-K_{S})_{0}$} & \colhead{$v_{\rm hel}$} & \colhead{e$v_{hel}$} & \colhead{S/N} & \colhead{EW1} & \colhead{EW2} & \colhead{UT} & \colhead{INST} & \colhead{class$^{a}$} & \colhead{Group$^{b}$} \\
 & \colhead{deg} & \colhead{deg} & \colhead{} & \colhead{} & \colhead{km s$^{-1}$} & \colhead{km s$^{-1}$} & \colhead{} & \colhead{} & \colhead{} & \colhead{} & \colhead{} & \colhead{}  & \\
}
\startdata
0000066+395650 &   112.36 &   -21.86 &    11.7 &    0.89 &    -48.3 &     4.2 &   20 &     4.8 &     5.3 &      2011-Nov-16 &   Goldcam &    D &    field \\
0004564+402930 &   113.45 &   -21.52 &    12.4 &    0.91 &    -33.3 &     2.3 &   12 &     3.6 &     3.9 &      2011-Nov-18 &   Goldcam &    D &    field \\
0006169+365203 &   112.97 &   -25.13 &    12.4 &    0.91 &    -49.1 &     3.7 &   31 &     3.1 &     3.6 &      2013-Oct-19 &   Modspec &    D &    field \\
0008452+235019 &   110.52 &   -38.00 &    11.6 &    0.92 &   -177.5 &     5.3 &   29 &     0.4 &     0.4 &      2011-Nov-11 &   Modspec &    G &  TriAnd1 \\
0010008+395243 &   114.36 &   -22.30 &    11.5 &    0.97 &   -128.0 &     5.1 &   27 &     0.5 &     0.8 &      2011-Nov-11 &   Modspec &    G &  TriAnd2 \\
0011284+293542 &   112.63 &   -32.48 &    11.3 &    0.92 &   -299.6 &     3.7 &   29 &     0.9 &     1.2 &      2013-Oct-19 &   Modspec &    G &    field \\
0011329+412258 &   114.94 &   -20.87 &    12.4 &    0.91 &     -3.5 &     5.2 &   22 &     2.1 &     2.4 &      2011-Nov-11 &   Modspec &    D &    field \\
0014415+371817 &   114.88 &   -24.99 &    10.9 &    0.99 &   -138.4 &     2.8 &   25 &     1.4 &     1.4 &      2012-Oct-29 &   Modspec &    G &  TriAnd1 \\
0015168+465156 &   116.53 &   -15.56 &    11.1 &    0.93 &   -160.9 &     3.4 &   36 &     1.0 &     1.2 &      2013-Oct-20 &   Modspec &    G &  TriAnd1 \\
0016023+450240 &   116.39 &   -17.37 &    11.9 &    0.87 &    -53.8 &    14.5 &   20 &     2.4 &     2.6 &      2011-Nov-16 &   Goldcam &    D &    field \\
0017402+462638 &   116.89 &   -16.03 &    11.3 &    0.94 &   -165.9 &     5.5 &   30 &     1.3 &     1.3 &      2011-Nov-18 &   Goldcam &    G &  TriAnd1 \\
0017434+235119 &   113.09 &   -38.37 &    12.1 &    0.92 &    -33.8 &     6.3 &   16 &     4.4 &     4.8 &      2011-Nov-19 &   Goldcam &    D &    field \\
0018381+235402 &   113.36 &   -38.36 &    12.2 &    0.91 &    -52.7 &     4.1 &   13 &     2.1 &     3.6 &      2011-Nov-19 &   Goldcam &    D &    field \\
0019276+421235 &   116.63 &   -20.27 &    11.7 &    0.92 &   -250.9 &     3.4 &   44 &     1.0 &     1.2 &      2013-Oct-20 &   Modspec &    G &    field \\
0020518+162645 &   112.39 &   -45.80 &    10.9 &    0.98 &   -179.9 &     3.6 &   40 &     0.8 &     1.2 &      2013-Oct-20 &   Modspec &    G &    field \\
0021538+455316 &   117.57 &   -16.68 &    11.5 &    0.95 &    -18.0 &     3.9 &   41 &     0.9 &     1.1 &      2013-Oct-19 &   Modspec &    G &    field \\
0022124+400956 &   116.89 &   -22.36 &    11.9 &    0.86 &    -26.0 &     1.2 &   16 &     3.0 &     3.7 &      2011-Nov-16 &   Goldcam &    D &    field \\
0022430+422802 &   117.30 &   -20.09 &    12.0 &    0.90 &   -149.7 &     5.3 &   30 &     0.6 &     0.8 &      2011-Nov-11 &   Modspec &    G &  TriAnd2 \\
0028312+401724 &   118.20 &   -22.37 &    11.3 &    0.94 &   -132.0 &     8.2 &   27 &     0.6 &     0.8 &      2011-Nov-19 &   Goldcam &    G &  TriAnd1 \\
0033083+234632 &   117.55 &   -38.90 &    12.0 &    0.86 &    -19.3 &     2.4 &   21 &     2.4 &     2.6 &      2011-Nov-17 &   Goldcam &    D &    field \\
0034189+272113 &   118.26 &   -35.36 &    11.8 &    0.87 &    -12.6 &     6.3 &   13 &     2.3 &     2.4 &      2011-Nov-16 &   Goldcam &    D &    field \\
0034442+460719 &   119.91 &   -16.65 &    10.9 &    0.99 &   -141.8 &     5.1 &   23 &     0.9 &     1.0 &      2012-Oct-28 &   Modspec &    G &  TriAnd1 \\
0036583+321110 &   119.38 &   -30.59 &    12.2 &    0.91 &      8.9 &     5.1 &   18 &     2.6 &     3.1 &      2011-Nov-15 &   Goldcam &    D &    field \\
0037165+334215 &   119.56 &   -29.08 &    12.4 &    0.90 &    -23.4 &     3.0 &   19 &     2.3 &     2.9 &      2012-Oct-28 &   Modspec &    D &    field \\
0037192+384457 &   119.92 &   -24.04 &    12.3 &    0.93 &    -11.8 &     1.9 &   14 &     3.9 &     4.6 &      2011-Nov-15 &   Goldcam &    D &    field \\
0038264+393424 &   120.21 &   -23.23 &    12.0 &    0.92 &    -19.8 &     9.8 &   22 &     4.7 &     5.0 &      2011-Nov-19 &   Goldcam &    D &    field \\
0038406+252313 &   119.30 &   -37.39 &    11.9 &    0.86 &     -5.6 &     3.8 &   22 &     3.3 &     3.9 &      2011-Nov-17 &   Goldcam &    D &    field \\
0039206+274800 &   119.67 &   -34.99 &    10.3 &    1.05 &   -291.4 &     2.1 &   34 &     0.8 &     1.1 &      2012-Oct-27 &   Modspec &    G &    field \\
0040357+384313 &   120.62 &   -24.10 &    10.8 &    1.00 &   -153.0 &     2.2 &   31 &     0.9 &     1.0 &      2012-Oct-27 &   Modspec &    G &  TriAnd1 \\
0040442+363805 &   120.54 &   -26.19 &    12.2 &    0.91 &    -45.3 &    10.0 &    2 &     8.3 &     9.8 &      2011-Nov-20 &   Goldcam &    D &    field \\
0042427+341227 &   120.87 &   -28.62 &    12.0 &    0.86 &    -11.9 &     3.2 &   21 &     3.4 &     3.7 &      2011-Nov-17 &   Goldcam &    D &    field \\
0043444+340830 &   121.12 &   -28.70 &    12.2 &    0.93 &    -21.2 &     4.9 &   24 &     2.8 &     3.4 &      2012-Oct-28 &   Modspec &    D &    field \\
0044567+273735 &   121.17 &   -35.22 &    12.0 &    0.90 &     -8.0 &     3.2 &   32 &     3.6 &     4.0 &      2012-Oct-28 &   Modspec &    D &    field \\
0047532+401734 &   122.20 &   -22.57 &    10.0 &    1.11 &   -132.0 &     1.9 &   28 &     1.1 &     1.2 &      2012-Oct-27 &   Modspec &    G &  TriAnd1 \\
0048394+422949 &   122.39 &   -20.37 &    10.6 &    1.01 &   -146.7 &     3.0 &   36 &     1.0 &     1.0 &      2012-Oct-29 &   Modspec &    G &  TriAnd1 \\
0050296+313838 &   122.70 &   -31.23 &    12.0 &    0.92 &    -34.3 &     1.8 &   26 &     3.4 &     3.6 &      2012-Oct-28 &   Modspec &    D &    field \\
0050308+362458 &   122.72 &   -26.46 &    12.3 &    0.91 &    -74.0 &     4.3 &   26 &     3.2 &     3.6 &      2012-Oct-28 &   Modspec &    D &    field \\
0051043+335407 &   122.84 &   -28.97 &    12.0 &    0.86 &     -2.7 &     3.2 &   23 &     1.7 &     1.9 &      2011-Nov-17 &   Goldcam &    G &    field \\
0051100+463939 &   122.88 &   -16.21 &    11.0 &    1.04 &   -178.5 &     3.7 &   34 &     0.6 &     0.8 &      2013-Oct-19 &   Modspec &    G &  TriAnd2 \\
0051470+280311 &   123.02 &   -34.81 &    11.8 &    0.87 &    -11.6 &     6.4 &   16 &     3.7 &     4.6 &      2011-Nov-16 &   Goldcam &    D &    field \\
0052055+393406 &   123.07 &   -23.30 &    12.4 &    0.91 &    -69.3 &     3.8 &   16 &     1.7 &     1.6 &      2011-Nov-18 &   Goldcam &    G &    field \\
0052304+393303 &   123.15 &   -23.32 &    10.7 &    1.02 &   -136.9 &     5.6 &   21 &     0.9 &     1.1 &      2011-Nov-18 &   Goldcam &    G &  TriAnd1 \\
0053235+395559 &   123.34 &   -22.94 &    11.0 &    0.96 &   -134.1 &     2.6 &   26 &     1.2 &     1.4 &      2012-Oct-29 &   Modspec &    G &  TriAnd1 \\
0054529+284300 &   123.84 &   -34.15 &    12.3 &    0.91 &    -34.6 &     6.1 &   15 &     3.3 &     3.8 &      2011-Nov-18 &   Goldcam &    D &    field \\
0054535+270317 &   123.88 &   -35.81 &    12.3 &    0.91 &    -15.3 &     5.5 &   15 &     3.0 &     3.8 &      2011-Nov-18 &   Goldcam &    D &    field \\
0055151+215224 &   124.10 &   -40.99 &    11.5 &    0.92 &     10.4 &     7.3 &   29 &     3.3 &     3.6 &      2011-Nov-18 &   Goldcam &    D &    field \\
0055369+205253 &   124.24 &   -41.97 &    11.8 &    0.88 &     -8.6 &     2.8 &   23 &     3.4 &     3.8 &      2011-Nov-17 &   Goldcam &    D &    field \\
0055509+411360 &   123.82 &   -21.63 &    12.2 &    0.90 &   -114.7 &     3.9 &   23 &     1.7 &     1.3 &      2013-Oct-19 &   Modspec &    G &  TriAnd2 \\
0055551+372146 &   123.92 &   -25.50 &    12.4 &    0.90 &    -18.4 &     4.8 &   18 &     3.4 &     3.6 &      2011-Nov-18 &   Goldcam &    D &    field \\
0056385+362637 &   124.10 &   -26.42 &    12.0 &    0.91 &    -35.8 &     4.7 &   16 &     3.6 &     3.5 &      2011-Nov-18 &   Goldcam &    D &    field \\
0056509+463713 &   123.90 &   -16.24 &    11.8 &    0.88 &    -36.2 &     3.2 &   18 &     2.5 &     2.9 &      2011-Nov-17 &   Goldcam &    D &    field \\
0058360+450047 &   124.26 &   -17.84 &    11.8 &    0.88 &    -71.2 &     3.1 &   20 &     0.7 &     0.7 &      2011-Nov-17 &   Goldcam &    G &  TriAnd1 \\
0059128+414245 &   124.48 &   -21.13 &    11.9 &    0.87 &    -13.0 &     2.2 &   14 &     3.2 &     3.6 &      2011-Nov-16 &   Goldcam &    D &    field \\
0059164+385602 &   124.60 &   -23.91 &    11.7 &    0.93 &   -119.9 &     1.9 &   22 &     0.5 &    -0.7 &      2011-Nov-12 &   Modspec &    G &  TriAnd2 \\
0101184+374507 &   125.09 &   -25.08 &    11.9 &    0.90 &    -45.0 &     2.5 &   12 &     2.4 &     3.0 &      2011-Nov-12 &   Modspec &    D &    field \\
0102438+143500 &   127.03 &   -48.20 &    12.2 &    0.90 &     -7.4 &     3.6 &   32 &     2.3 &     2.7 &      2013-Oct-20 &   Modspec &    D &    field \\
0103574+284503 &   126.24 &   -34.04 &    12.2 &    0.92 &    -25.9 &     3.7 &   28 &     3.6 &     4.1 &      2013-Oct-19 &   Modspec &    D &    field \\
0104077+400235 &   125.56 &   -22.76 &    12.0 &    0.86 &    -30.4 &     1.5 &   24 &     3.4 &     3.7 &      2011-Nov-17 &   Goldcam &    D &    field \\
0104295+451314 &   125.34 &   -17.59 &    11.8 &    0.87 &   -114.8 &     3.9 &   26 &     0.6 &     0.6 &      2012-Oct-28 &   Modspec &    G &  TriAnd1 \\
0106121+414018 &   125.89 &   -21.12 &     9.8 &    1.10 &   -187.2 &     2.7 &   26 &     1.2 &     1.3 &      2012-Oct-29 &   Modspec &    G &  TriAnd1 \\
0108205+381304 &   126.58 &   -24.53 &    10.9 &    1.02 &   -117.0 &     1.9 &   19 &     1.5 &     1.5 &      2012-Oct-29 &   Modspec &    G &  TriAnd1 \\
0108513+224421 &   128.17 &   -39.96 &    12.0 &    0.91 &      9.6 &     6.3 &   13 &     3.2 &     3.8 &      2011-Nov-18 &   Goldcam &    D &    field \\
0109330+392120 &   126.75 &   -23.38 &    10.2 &    1.08 &   -128.2 &     2.0 &   28 &     0.7 &     0.3 &      2012-Oct-27 &   Modspec &    G &  TriAnd1 \\
0112312+202928 &   129.59 &   -42.10 &    11.8 &    0.88 &    -29.3 &     5.1 &   17 &     2.8 &     0.2 &      2011-Nov-17 &   Goldcam &    G &  TriAnd1 \\
0114318+400412 &   127.71 &   -22.59 &    12.1 &    0.90 &    -35.1 &    10.0 &    9 &     3.5 &     3.7 &      2011-Nov-20 &   Goldcam &    D &    field \\
0118006+301415 &   127.05 &   -24.32 &    10.8 &    0.97 &    -99.8 &     5.3 &   28 &     1.1 &     1.3 &      2011-Nov-11 &   Modspec &    G &  TriAnd1 \\
0118151+414947 &   128.27 &   -20.76 &    11.7 &    0.93 &   -132.2 &     4.1 &   23 &     0.8 &     0.8 &      2012-Oct-28 &   Modspec &    G &  TriAnd2 \\
0118518+422120 &   128.33 &   -20.23 &     9.9 &    1.09 &   -168.7 &     4.9 &   26 &     0.7 &     1.0 &      2012-Oct-28 &   Modspec &    G &  TriAnd1 \\
0120563+410332 &   128.90 &   -21.47 &    10.8 &    1.00 &   -152.1 &     4.7 &   27 &     1.0 &     1.4 &      2012-Oct-28 &   Modspec &    G &  TriAnd1 \\
0123174+335858 &   130.44 &   -28.42 &    11.4 &    0.94 &   -111.9 &     3.3 &   26 &     0.6 &     1.1 &      2013-Oct-20 &   Modspec &    G &  TriAnd1 \\
0125025+223446 &   133.02 &   -39.63 &    11.9 &    0.91 &    -43.9 &     3.8 &   24 &     3.5 &     4.0 &      2012-Oct-28 &   Modspec &    D &    field \\
0126456+211847 &   133.82 &   -40.81 &    11.9 &    0.86 &      5.1 &     8.0 &   21 &     4.0 &     4.0 &      2011-Nov-16 &   Goldcam &    D &    field \\
0129316+304749 &   132.51 &   -31.36 &    11.8 &    0.87 &    -44.2 &     2.4 &   22 &     2.7 &     3.3 &      2011-Nov-17 &   Goldcam &    D &    field \\
0129510+421019 &   130.49 &   -20.13 &    11.9 &    0.87 &     -9.5 &     5.4 &   21 &     4.3 &     4.5 &      2011-Nov-18 &   Goldcam &    D &    field \\
0131068+454442 &   130.13 &   -16.57 &    10.6 &    1.02 &   -148.9 &     5.0 &   10 &     0.6 &     0.8 &      2011-Nov-11 &   Modspec &    G &  TriAnd1 \\
0131323+351934 &   132.09 &   -26.83 &    11.9 &    0.92 &    -21.1 &     5.3 &   10 &     3.9 &     4.2 &      2011-Nov-18 &   Goldcam &    D &    field \\
0131447+331853 &   132.54 &   -28.80 &    11.1 &    0.95 &   -231.6 &     4.0 &   36 &     0.4 &     0.5 &      2012-Oct-28 &   Modspec &    G &    field \\
0135157+392625 &   132.08 &   -22.65 &    10.9 &    0.94 &   -134.0 &     3.4 &   32 &     1.1 &     1.2 &      2013-Oct-20 &   Modspec &    G &  TriAnd1 \\
0135572+445814 &   131.15 &   -17.19 &    10.6 &    1.04 &   -115.6 &     3.7 &   25 &     0.9 &     1.0 &      2013-Oct-19 &   Modspec &    G &  TriAnd1 \\
0138005+310604 &   134.53 &   -30.72 &    11.8 &    0.93 &   -138.6 &     5.5 &   20 &     1.5 &     1.7 &      2011-Nov-17 &   Goldcam &    G &  TriAnd2 \\
0143280+395547 &   133.65 &   -21.86 &    11.3 &    0.95 &    -95.5 &     2.1 &   30 &     0.8 &     0.8 &      2012-Oct-27 &   Modspec &    G &  TriAnd1 \\
0145176+291850 &   136.80 &   -32.10 &    11.8 &    0.92 &    -30.0 &     4.0 &   19 &     4.5 &     4.7 &      2011-Nov-15 &   Goldcam &    D &    field \\
0145338+284751 &   137.02 &   -32.59 &    10.9 &    0.96 &    -91.5 &     1.7 &   30 &     0.5 &     0.8 &      2012-Oct-27 &   Modspec &    G &  TriAnd1 \\
0146203+360440 &   135.19 &   -25.48 &    10.5 &    1.03 &   -102.8 &    10.0 &   17 &     1.1 &     1.1 &      2011-Nov-20 &   Goldcam &    G &  TriAnd1 \\
0146216+402605 &   134.11 &   -21.24 &    12.3 &    0.91 &    -35.3 &     3.3 &   25 &     3.8 &     4.3 &      2013-Oct-20 &   Modspec &    D &    field \\
0146291+371501 &   134.93 &   -24.34 &    10.9 &    0.96 &   -102.0 &     2.9 &   28 &     0.8 &     1.0 &      2012-Oct-29 &   Modspec &    G &  TriAnd1 \\
0146402+440851 &   133.28 &   -17.61 &    11.9 &    0.86 &   -109.3 &     5.1 &   18 &     0.5 &     0.7 &      2011-Nov-18 &   Goldcam &    G &  TriAnd1 \\
0150346+364724 &   135.92 &   -24.59 &    11.8 &    0.91 &    -11.0 &     1.1 &    9 &     3.1 &     3.8 &      2011-Nov-17 &   Goldcam &    D &    field \\
0152538+394641 &   135.58 &   -21.57 &    11.9 &    0.87 &      3.7 &     2.2 &   19 &     3.2 &     3.4 &      2011-Nov-15 &   Goldcam &    D &    field \\
0154085+382029 &   136.23 &   -22.91 &    10.5 &    1.01 &    -63.3 &     1.6 &   22 &     0.3 &     0.4 &      2012-Oct-29 &   Modspec &    G &  TriAnd1 \\
0154522+361326 &   137.00 &   -24.91 &    10.4 &    1.02 &    -75.3 &     1.8 &   29 &     0.8 &     0.8 &      2012-Oct-29 &   Modspec &    G &  TriAnd1 \\
0155244+410614 &   135.71 &   -20.18 &    10.4 &    1.03 &    -99.9 &     1.5 &   30 &     0.9 &     1.3 &      2012-Oct-29 &   Modspec &    G &  TriAnd1 \\
0155575+392150 &   136.31 &   -21.83 &    11.3 &    0.95 &   -126.7 &     3.4 &   33 &     1.1 &     1.2 &      2013-Oct-20 &   Modspec &    G &  TriAnd1 \\
0158234+461103 &   134.85 &   -15.13 &    10.9 &    0.97 &    -82.2 &     3.4 &   31 &     1.0 &     1.0 &      2013-Oct-20 &   Modspec &    G &  TriAnd1 \\
0159430+344015 &   138.53 &   -26.12 &    10.4 &    1.07 &   -127.2 &     2.3 &   28 &     0.9 &     0.9 &      2012-Oct-27 &   Modspec &    G &  TriAnd1 \\
0204068+305222 &   140.83 &   -29.46 &    10.5 &    1.02 &    -89.5 &     1.2 &   24 &     1.4 &     1.4 &      2012-Oct-29 &   Modspec &    G &  TriAnd1 \\
0205311+443659 &   136.56 &   -16.29 &    11.1 &    0.99 &   -101.5 &     3.7 &   30 &     1.3 &     1.3 &      2013-Oct-19 &   Modspec &    G &  TriAnd1 \\
0206248+092837 &   151.90 &   -49.13 &    11.9 &    0.96 &   -188.3 &     2.1 &   29 &     0.9 &     1.2 &      2012-Oct-27 &   Modspec &    G &    field \\
0208047+433706 &   137.33 &   -17.10 &    11.1 &    0.97 &   -122.0 &     3.3 &   18 &     1.2 &     1.4 &      2013-Oct-20 &   Modspec &    G &  TriAnd1 \\
0208241+343703 &   140.44 &   -25.62 &    10.0 &    1.11 &    -90.7 &     1.8 &   31 &     1.1 &     1.3 &      2012-Oct-27 &   Modspec &    G &  TriAnd1 \\
0210532+354315 &   140.57 &   -24.42 &    10.3 &    1.07 &    -95.8 &     2.2 &   32 &     1.1 &     1.2 &      2012-Oct-27 &   Modspec &    G &  TriAnd1 \\
0215010+391644 &   140.11 &   -20.79 &     9.5 &    1.13 &    -89.2 &     2.2 &   27 &     1.3 &     1.3 &      2012-Oct-27 &   Modspec &    G &  TriAnd1 \\
0216479+181401 &   149.81 &   -40.14 &    10.9 &    1.02 &   -264.7 &     3.8 &   23 &     0.9 &     0.8 &      2012-Oct-29 &   Modspec &    G &    field \\
0218301+145038 &   152.29 &   -43.01 &    11.4 &    1.00 &   -232.1 &     2.1 &   16 &     0.8 &     1.1 &      2012-Oct-29 &   Modspec &    G &    field \\
0222450+061341 &   159.80 &   -49.99 &    12.2 &    0.93 &   -178.2 &     3.3 &   27 &     1.1 &     1.3 &      2013-Oct-20 &   Modspec &    G &    field \\
0230089+293925 &   147.32 &   -28.50 &    11.2 &    0.96 &    -87.1 &     3.6 &   27 &     1.2 &     1.3 &      2012-Oct-28 &   Modspec &    G &  TriAnd1 \\
0232082+311429 &   147.00 &   -26.88 &    10.4 &    1.03 &    -60.8 &     3.9 &   24 &     1.4 &     1.4 &      2012-Oct-28 &   Modspec &    G &  TriAnd1 \\
0232283+430640 &   141.84 &   -16.02 &    11.9 &    0.93 &   -130.6 &     3.5 &   30 &     0.8 &     0.9 &      2013-Oct-20 &   Modspec &    G &  TriAnd2 \\
0235400+362908 &   145.29 &   -21.82 &    10.3 &    1.04 &   -100.1 &     4.3 &   22 &     1.4 &     1.6 &      2012-Oct-28 &   Modspec &    G &  TriAnd1 \\
0237250+370342 &   145.37 &   -21.15 &    12.3 &    0.91 &    -29.9 &     3.8 &   31 &     3.1 &     3.5 &      2013-Oct-20 &   Modspec &    D &    field \\
0238059+172905 &   156.05 &   -38.43 &    11.8 &    0.97 &   -211.0 &     2.3 &   16 &     0.6 &     0.8 &      2012-Oct-29 &   Modspec &    G &    field \\
0240181+375140 &   145.56 &   -20.18 &    10.7 &    1.00 &    -98.8 &     4.0 &   21 &     1.4 &     1.4 &      2012-Oct-28 &   Modspec &    G &  TriAnd1 \\
0242433+283308 &   150.67 &   -28.26 &    10.5 &    1.07 &    -65.7 &     1.8 &   28 &     1.0 &     1.3 &      2012-Oct-27 &   Modspec &    G &  TriAnd1 \\
0243131+332942 &   148.22 &   -23.83 &     9.5 &    1.14 &    -61.5 &     1.9 &   25 &     0.5 &     0.7 &      2012-Oct-27 &   Modspec &    G &  TriAnd1 \\
0248112+360447 &   147.92 &   -21.06 &    10.0 &    1.09 &    -54.1 &     1.9 &   27 &     0.7 &     0.9 &      2012-Oct-27 &   Modspec &    G &  TriAnd1 \\
0248197+380710 &   146.94 &   -19.24 &    11.6 &    0.93 &    -75.2 &     3.7 &   38 &     0.9 &     0.9 &      2013-Oct-20 &   Modspec &    G &  TriAnd1 \\
0251063+165335 &   159.78 &   -37.27 &    11.3 &    1.04 &   -197.5 &     3.8 &    6 &     1.1 &     1.5 &      2012-Oct-29 &   Modspec &    G &    field \\
0252502+312423 &   151.27 &   -24.70 &    10.1 &    1.08 &    -54.9 &     3.7 &   25 &     0.7 &     0.8 &      2013-Oct-19 &   Modspec &    G &  TriAnd1 \\
0253000+375204 &   147.92 &   -19.03 &    11.4 &    0.99 &   -112.3 &     3.8 &   23 &     0.9 &     1.2 &      2013-Oct-20 &   Modspec &    G &  TriAnd2 \\
0254132+395433 &   147.12 &   -17.13 &    11.5 &    0.92 &    -81.0 &     3.8 &   42 &     0.7 &     0.8 &      2013-Oct-19 &   Modspec &    G &  TriAnd1 \\
0259473+374837 &   149.19 &   -18.43 &    11.0 &    0.94 &    -82.3 &     3.7 &   39 &     0.9 &     0.9 &      2013-Oct-19 &   Modspec &    G &  TriAnd1 \\
0300140+263657 &   155.57 &   -27.97 &    10.3 &    1.07 &    -87.8 &     1.6 &   19 &     1.7 &     1.7 &      2012-Oct-29 &   Modspec &    G &  TriAnd1 \\
0305474+271010 &   156.41 &   -26.83 &    11.3 &    0.91 &    -68.4 &     3.1 &   29 &     1.0 &     1.2 &      2013-Oct-20 &   Modspec &    G &  TriAnd1 \\
0306248+331250 &   152.94 &   -21.68 &    11.4 &    0.95 &   -125.6 &     3.7 &   27 &     1.2 &     1.3 &      2013-Oct-19 &   Modspec &    G &  TriAnd1 \\
0306391+252506 &   157.69 &   -28.18 &    11.7 &    0.96 &    -55.3 &     1.1 &   26 &     0.5 &     0.5 &      2012-Oct-27 &   Modspec &    G &  TriAnd2 \\
0307189+263423 &   157.09 &   -27.14 &    10.6 &    1.02 &    -95.2 &     2.5 &   16 &     0.7 &     0.8 &      2012-Oct-29 &   Modspec &    G &  TriAnd1 \\
0307331+350656 &   152.07 &   -19.94 &    10.1 &    1.08 &    -61.1 &     3.7 &   26 &     1.2 &     1.4 &      2013-Oct-19 &   Modspec &    G &  TriAnd1 \\
0311305+325132 &   154.11 &   -21.40 &    10.5 &    1.02 &   -119.6 &     3.6 &   28 &     0.9 &     1.0 &      2013-Oct-19 &   Modspec &    G &  TriAnd1 \\
0314314+243419 &   159.89 &   -27.85 &    11.4 &    1.03 &   -269.9 &     2.3 &   30 &     1.1 &     1.1 &      2012-Oct-27 &   Modspec &    G &    field \\
0318211+315433 &   155.97 &   -21.38 &    11.3 &    0.92 &    -36.9 &     3.6 &   22 &     0.9 &     1.3 &      2013-Oct-19 &   Modspec &    G &  TriAnd1 \\
0319503+264652 &   159.51 &   -25.36 &    10.0 &    1.11 &    -83.5 &     1.8 &   27 &     0.8 &     0.7 &      2012-Oct-27 &   Modspec &    G &  TriAnd1 \\
2254402+421852 &   100.95 &   -15.52 &    11.0 &    0.94 &   -118.4 &     3.7 &   30 &     1.3 &     1.4 &      2013-Oct-19 &   Modspec &    G &  TriAnd1 \\
2259157+423030 &   101.82 &   -15.72 &    11.9 &    0.86 &    -53.6 &     2.7 &   25 &     0.7 &     0.7 &      2011-Nov-17 &   Goldcam &    G &    field \\
2309342+375615 &   101.67 &   -20.69 &    11.9 &    0.86 &    -20.1 &     1.2 &   16 &     4.3 &     4.7 &      2011-Nov-16 &   Goldcam &    D &    field \\
2311295+415714 &   103.74 &   -17.16 &    11.8 &    0.92 &   -204.0 &     3.7 &   30 &     0.9 &     1.0 &      2013-Oct-19 &   Modspec &    G &  TriAnd2 \\
2312367+355340 &   101.38 &   -22.81 &    12.4 &    0.91 &    -75.7 &    39.8 &    6 &     1.2 &     1.6 &      2011-Nov-18 &   Goldcam &    G &    field \\
2314133+332953 &   100.65 &   -25.13 &    10.9 &    0.96 &   -201.4 &     2.5 &   28 &     0.4 &     0.5 &      2012-Oct-29 &   Modspec &    G &  TriAnd1 \\
2314262+412531 &   104.05 &   -17.86 &    11.8 &    0.99 &   -157.4 &     3.4 &   30 &     1.2 &     1.5 &      2013-Oct-20 &   Modspec &    G &  TriAnd2 \\
2314377+440800 &   105.16 &   -15.37 &    10.2 &    1.05 &   -180.8 &     4.8 &   13 &     0.4 &     0.8 &      2011-Nov-11 &   Modspec &    G &  TriAnd1 \\
2314420+371133 &   102.36 &   -21.78 &    10.4 &    1.04 &   -172.6 &     2.5 &   27 &     1.0 &     1.1 &      2012-Oct-27 &   Modspec &    G &  TriAnd1 \\
2317294+370149 &   102.84 &   -22.14 &    11.2 &    0.93 &   -312.9 &     5.4 &   33 &     0.7 &     0.9 &      2011-Nov-11 &   Modspec &    G &    field \\
2317414+311304 &   100.38 &   -27.52 &    10.7 &    1.00 &    -84.5 &     2.1 &   28 &     0.5 &     0.8 &      2012-Oct-27 &   Modspec &    G &  TriAnd1 \\
2318119+435650 &   105.71 &   -15.78 &    10.3 &    1.04 &   -196.3 &     5.5 &   18 &     0.5 &     0.8 &      2011-Nov-18 &   Goldcam &    G &  TriAnd1 \\
2318337+343707 &   102.05 &   -24.46 &     9.9 &    1.12 &   -203.9 &     2.3 &   33 &     0.9 &     1.0 &      2012-Oct-27 &   Modspec &    G &  TriAnd1 \\
2318590+372440 &   103.29 &   -21.91 &    11.2 &    0.96 &   -143.9 &     2.1 &   24 &     0.5 &     0.8 &      2012-Oct-27 &   Modspec &    G &  TriAnd1 \\
2322513+353207 &   103.32 &   -23.94 &    11.9 &    0.90 &      9.5 &     1.6 &    8 &     1.4 &     1.3 &      2011-Nov-15 &   Goldcam &    G &    field \\
2327114+422439 &   106.75 &   -17.79 &    11.8 &    0.88 &    -49.9 &     2.9 &   13 &     4.2 &     4.8 &      2011-Nov-16 &   Goldcam &    D &    field \\
2328209+400142 &   106.12 &   -20.11 &    11.8 &    0.88 &    -27.3 &     1.6 &   10 &     2.9 &     3.6 &      2011-Nov-16 &   Goldcam &    D &    field \\
2332500+404525 &   107.23 &   -19.71 &    11.1 &    1.06 &   -170.9 &     3.3 &   34 &     0.9 &     1.0 &      2013-Oct-20 &   Modspec &    G &  TriAnd2 \\
2333505+450925 &   108.86 &   -15.58 &    11.4 &    0.90 &   -138.8 &     3.7 &   32 &     0.7 &     0.8 &      2013-Oct-19 &   Modspec &    G &  TriAnd1 \\
2334021+235834 &   101.07 &   -35.59 &    12.1 &    0.94 &    -32.0 &     2.9 &   24 &     3.6 &     4.2 &      2012-Oct-28 &   Modspec &    D &    field \\
2337246+395922 &   107.86 &   -20.71 &    11.8 &    0.98 &   -172.6 &     5.3 &   16 &     0.7 &     1.0 &      2011-Nov-11 &   Modspec &    G &  TriAnd2 \\
2338336+400047 &   108.09 &   -20.75 &    10.6 &    1.04 &   -145.8 &     2.4 &   25 &     1.3 &     1.4 &      2012-Oct-29 &   Modspec &    G &  TriAnd1 \\
2339513+393604 &   108.21 &   -21.21 &    11.7 &    0.89 &    -25.7 &     2.3 &   19 &     2.7 &     2.9 &      2011-Nov-16 &   Goldcam &    D &    field \\
2345376+352056 &   108.09 &   -25.62 &    11.7 &    0.98 &   -142.7 &     2.4 &   42 &     1.1 &     1.1 &      2012-Oct-27 &   Modspec &    G &  TriAnd2 \\
2346094+455731 &   111.24 &   -15.43 &    11.9 &    0.91 &   -156.6 &     3.6 &   39 &     1.0 &     1.1 &      2013-Oct-19 &   Modspec &    G &  TriAnd2 \\
2346386+214809 &   103.57 &   -38.62 &    10.2 &    1.08 &   -263.1 &     1.9 &   35 &     0.6 &     0.9 &      2012-Oct-29 &   Modspec &    G &    field \\
2346533+223942 &   103.98 &   -37.82 &    12.4 &    0.90 &    -23.6 &     5.6 &   14 &     3.2 &     3.1 &      2011-Nov-18 &   Goldcam &    D &    field \\
2347422+235641 &   104.71 &   -36.66 &    12.5 &    0.90 &    -43.7 &     2.6 &    8 &     3.0 &     3.4 &      2012-Oct-28 &   Modspec &    D &    field \\
2347568+420218 &   110.50 &   -19.29 &    11.8 &    0.89 &   -337.9 &     4.3 &   27 &     0.6 &     0.9 &      2011-Nov-17 &   Goldcam &    G &    field \\
2348498+454925 &   111.67 &   -15.67 &     9.8 &    1.10 &   -112.0 &     3.3 &   22 &     1.0 &     1.5 &      2013-Oct-20 &   Modspec &    G &  TriAnd1 \\
2349423+300654 &   107.37 &   -30.88 &    11.8 &    0.95 &   -187.2 &     3.4 &   34 &     0.6 &     0.6 &      2013-Oct-20 &   Modspec &    G &  TriAnd2 \\
2350273+402510 &   110.55 &   -20.98 &    12.4 &    0.91 &    -37.2 &     4.9 &   15 &     3.3 &     3.8 &      2011-Nov-15 &   Goldcam &    D &    field \\
2350566+415721 &   111.05 &   -19.51 &    11.7 &    0.89 &    -38.4 &     1.9 &   16 &     4.0 &     4.3 &      2011-Nov-16 &   Goldcam &    D &    field \\
2352223+223401 &   105.48 &   -38.29 &    11.9 &    0.86 &    -15.0 &     3.5 &   14 &     4.1 &     4.0 &      2011-Nov-17 &   Goldcam &    D &    field \\
2353556+335644 &   109.51 &   -27.42 &    12.3 &    0.91 &     -8.3 &     3.9 &   24 &     2.2 &     2.7 &      2013-Oct-19 &   Modspec &    D &    field \\
2355440+290121 &   108.52 &   -32.29 &    10.6 &    0.99 &   -148.9 &     3.7 &   24 &     1.0 &     1.2 &      2013-Oct-19 &   Modspec &    G &  TriAnd1 \\
2356283+210505 &   106.11 &   -39.98 &    11.8 &    0.88 &     -3.5 &     5.0 &   11 &     3.6 &     3.9 &      2011-Nov-16 &   Goldcam &    D &    field \\
2358446+355360 &   111.10 &   -25.76 &    10.8 &    1.03 &   -117.3 &     1.9 &   36 &     0.9 &     0.9 &      2012-Oct-27 &   Modspec &    G &  TriAnd1 \\
2358463+120822 &   103.21 &   -48.71 &    11.5 &    0.91 &    -83.0 &     3.8 &   31 &     0.8 &     0.9 &      2013-Oct-19 &   Modspec &    G &  TriAnd1 \\
\enddata

\begin{enumerate}[a]
\item Class refers to either dwarf (D) or giant (G) stars.
\item Group indicates which stars have been identified as members of either TriAnd1 or TriAnd2. All others are labeled ``field'' for field stars.
\end{enumerate}
\end{deluxetable}

\begin{deluxetable}{c c c c c c c c c c c c c c}
\rotate
\tabletypesize{\scriptsize}
\tablewidth{0pt}
\tablecaption{\label{dups}Stars with repeat observations.}
\tablehead{
\colhead{ID} & \colhead{$v_{\rm hel}$} & \colhead{UT} & \colhead{INST$^{c}$} & \colhead{$\sigma$(RV)} \\
 & \colhead{km s$^{-1}$} &  &  &  \colhead{km s$^{-1}$} \\
}
\startdata
0019276+421235	&	-248.7	&	2012-Oct-29	&	Modspec	&	1.6	\\
		&	-250.9	&	2013-Oct-20	&	Modspec	&		\\
0025311+382840	&	6.5	&	2011-Nov-11	&	Modspec	&	1.1	\\
		&	8.0	&			&	RP04	&		\\
0028291+375339	&	-280.0	&	2011-Nov-19	&	Goldcam	&	24.7	\\
		&	-245.1	&			&	RP04	&		\\
0037344+404243	&	-229.3	&	2011-Nov-19	&	Goldcam	&	13.4	\\
		&	-210.4	&			&	RP04	&		\\
0052304+393303	&	-133.4	&	2012-Oct-29	&	Modspec	&	2.5	\\
		&	-136.9	&	2011-Nov-18	&	Goldcam	&		\\
0054518+351060	&	-141.6	&	2011-Nov-18	&	Goldcam	&	7.6	\\
		&	-130.9	&			&	RP04	&		\\
0100230+295724	&	-95.9	&	2011-Nov-12	&	Modspec	&	5.9	\\
		&	-107.7	&			&	RP04	&		\\
		&	-101.0	&	2012-Oct-27	&	Modspec	&		\\
0111039+355319	&	-135.6	&	2012-Oct-27	&	Modspec	&	3.5	\\
		&	-140.5	&			&	RP04	&		\\
0121416+363550	&	-103.1	&	2011-Nov-15	&	Goldcam	&	13.0	\\
		&	-121.5	&			&	RP04	&		\\
0137297+371938	&	-141.6	&	2011-Nov-18	&	Goldcam	&	5.9	\\
		&	-133.2	&	 		&	RP04	&		\\
0140365+364907	&	-111.9	&	2011-Nov-15	&	Goldcam	&	3.5	\\
		&	-107.0	&			&	RP04	&		\\
0142564+385120	&	-126.9	&	2011-Nov-16	&	Goldcam	&	0.7	\\
		&	-127.9	&			&	RP04	&		\\
0143280+395547	&	-108.3	&	2011-Nov-15	&	Goldcam	&	9.1	\\
		&	-95.5	&	2012-Oct-27	&	Modspec	&		\\
0146203+360440	&	-111.0	&	2011-Nov-12	&	Modspec	&	8.3	\\
		&	-102.8	&	2011-Nov-20	&	Goldcam	&		\\
		&	-94.5	&	2012-Oct-29	&	Modspec	&		\\
2333383+390924	&	-166.9	&	2011-Nov-15	&	Goldcam	&	12.1	\\
		&	-180.8	&			&	RP04	&		\\
		&	-190.9	&	2012-Oct-29	&	Modspec	&		\\
2337246+395922	&	-161.5	&	2011-Nov-15	&	Goldcam	&	7.8	\\
		&	-172.6	&	2011-Nov-11	&	Modspec	&		\\
2338251+392746	&	-187.0	&	2011-Nov-15	&	Goldcam	&	3.8	\\
		&	-181.6	&			&	RP04	&		\\
2349054+405731	&	-159.2	&	2011-Nov-18	&	Goldcam	&	8.3	\\
		&	-145.4	&			&	RP04	&		\\
		&	-160.2	&	2012-Oct-29	&	Modspec	&		\\
2358348+300936	&	-183.1	&	2012-Oct-27	&	Modspec	&	3.3	\\
		&	-178.5	&			&	RP04	&		\\
\enddata

\begin{enumerate}[c]
\item Stars labeled as ``RP04'' were observed by Rocha-Pinto et al. 2004 using the B \& C Spectrograph on the Bok 2.3-m telescope at Steward Observatory.
\end{enumerate}
\end{deluxetable}

\end{document}